\def\simlt{\mathrel{\rlap{\lower 3pt\hbox{$\sim$}}\raise
2.0pt\hbox{$<$}}}
\def\simgt{\mathrel{\rlap{\lower 3pt\hbox{$\sim$}} \raise
2.0pt\hbox{$>$}}}

\newcommand{\q}{\begin{equation}}
\newcommand{\qa}{\begin{eqnarray}}
\newcommand{\qs}{\begin{eqnarray*}}
\newcommand{\nq}{\end{equation}}
\newcommand{\nqa}{\end{eqnarray}}
\newcommand{\nqs}{\end{eqnarray*}}
\def\be{\begin{equation}}
\def\ee{\end{equation}}

\documentclass[useAMS,usegraphicx]{mn2e}

\def\aj{AJ }
\def\apj{ApJ }

\newcommand{\gsim}{\raisebox{-3.8pt}{$\;\stackrel{\textstyle >}{\sim}\;$}}
\newcommand{\lsim}{\raisebox{-3.8pt}{$\;\stackrel{\textstyle <}{\sim}\;$}}

\title[Evolution of Red Sequence]{The evolution of the galaxy Red Sequence in simulated clusters and groups}
\author[Romeo et al.]{
A.D. Romeo$^{1,2}$ \thanks{E-mail: aro@na.astro.it},
N.R. Napolitano$^{1}$,
G. Covone$^{1,5}$,
J. Sommer-Larsen$^{3,4}$, \and V. Antonuccio-Delogu$^{2,6}$\thanks{Marie Curie senior fellow}, 
M. Capaccioli$^{5}$\\ \\
$^{1}$ INAF--Osservatorio Astronomico di Capodimonte, v. Moiariello 16, 80131 Naples, Italy\\
$^{2}$ INAF--Osservatorio Astrofisico di Catania, v. S.Sofia 78, 95123 Catania, Italy\\
$^{3}$ Dark Cosmology Centre, Juliane Maries Vej 30, 2100 Copenhagen \O, Denmark\\
$^{4}$ Institute of Astronomy, University of Tokyo, Osawa 2-21-1, Mitaka, Tokyo, 181-0015, Japan\\
$^{5}$ Dipartimento di Scienze Fisiche, Universit\`a Federico II, Via Cinthia, 80126, Napoli, Italy\\
$^{6}$ Astrophysics, Department of Physics, University of Oxford, Keble Road, Oxford OX1 3RH, UK
}

\begin{document}

\date{Accepted ?. Received ?; in original form ?}

\pagerange{\pageref{firstpage}--\pageref{lastpage}} \pubyear{2006}

\maketitle

\label{firstpage}
\begin{abstract}
N--body + hydrodynamical simulations of the formation and evolution
of galaxy groups and clusters in a $\Lambda$CDM cosmology are used in order to follow the 
building-up of the colour-magnitude relation in two clusters and in 12 groups.
We have found that galaxies, starting from the more massive, move to the Red Sequence (RS)
as they get aged over times and eventually set upon a ``dead sequence'' (DS)
once they have stopped their bulk star formation
activity. Fainter galaxies keep having significant star formation
out to very recent epochs and lie broader around the RS.
Environment plays a role as
galaxies in groups and cluster outskirts hold star formation activity
longer than the central cluster regions. However galaxies 
experiencing infall from the outskirts to the central parts keep star
formation on until they settle on to the DS of the core galaxies.
Merging contributes to mass assembly until $z\sim 1$, after which major events
only involve the brightest cluster galaxies. 

The emerging scenario is that 
the evolution of the colour-magnitude properties of galaxies within the hierarchical framework
is mainly driven by star formation activity during dark matter halos
assembly. Galaxies progressively quenching their star formation settle to a very sharp 
``red and dead'' sequence, which
turns out to be universal, its slope and scatter being almost independent 
of the redshift (since at least z$\sim1.5$) and environment. 

Differently from the DS, the operatively defined RS evolves more evidently with $z$,
the epoch when it changes its slope being closely corresponding to that
at which the passive galaxies population takes over the star forming one: this goes
from $z\simeq$1 in clusters down to 0.4 in normal groups.


\end{abstract}


\section{Introduction}

Within the standard $\Lambda$CDM cosmological model, dark matter (DM) haloes 
evolve through hierarchical assembly to form galaxies; their gravitational-only dynamics 
is powerfully described by N-body (collisionless) simulations
(since e.g. Evrard 1990, Thomas \& Couchman 1992, Navarro, Frenk \& White 1995).
Yet, galaxy properties are determined by the interplay between DM itself and the baryons
infalling and settling in to the DM potential (White \& Rees 1978). 
Therefore, a fully physical
approach has to include all the mechanisms driving the baryons' evolution 
within DM halos: 
radiative cooling of the gas, star formation, and the subsequent
feedback from the latter on to the gas again, in form of both chemical 
enrichment and energy release from supernovae (SN), and possibly AGN feedbacks. 
Once traced the merging history
of the DM haloes in N-body simulations, a semi-analytical method is a useful
means for modelling the galaxies and their stellar populations, by incorporating
``by hand'' a number of schemes, each implementing a physical module
(e.g. De Lucia, Kauffmann \& White, 2004a ). 
An alternative self-consistent approach
is that of re-simulating a smaller region, like a galaxy cluster, taken from the cosmological
simulation, adding the baryon physics via hydrodynamical codes: however, this can be attained
at expenses of a much higher computational cost, that depends on the mass and
spatial resolution (see Tornatore et al. 2004, Romeo et al. 2006). 
%
%


From the observational point of view, there are evidences of a so-called ``downsizing'' picture of the 
galaxy formation, both in the star formation process and the stellar mass assembly:
the former is found to proceed from more ($>10^{11}M_{\odot}$) to less ($<10^{10}M_{\odot}$) massive galaxies,
which translates in the bulk of stellar mass of bright systems being already in place at $z>$1
(Borch et al., 2006). This is mirrored by the typical mass of star-forming galaxies decreasing 
over time (Juneau et al. 2005, Bundy et al. 2006), or by the smooth decline of their characteristic 
$L^*$ (Cowie et al. 1996): the local and $z<$1 galaxy LF is highly populated at faint magnitudes 
and is shaped on the other side by a bright-end cutoff.
In terms of stellar mass, the lack of evolution of the bright end of the mass function since $z\sim 1$,
combined with the deficit of objects less massive than $10^{10}M_{\odot}$ (corresponding to $M^*$+2)
at $z\simeq$1 (Kodama et al. 2004, Fontana et al. 2006), give rise to a picture in which
massive galaxies are dominated by old stellar populations, mostly formed as starbursts across
an epoch ranging from $z\sim$4 to $\sim 1.5$, followed by milder evolution thereafter --whereas
less massive galaxies have more extended activity during the last 8 Gyrs.

Physical mechanisms responsible for downsizing root in correlation with local galaxy density,
in that quenching of star formation can be strongly driven by environment (Bundy et al. 2006,
Sheth et al. 2006, Thomas et al. 2005): 
star formation, hence mass assembly and chemical enrichment as well, occur at accelerated pace
in massive galaxies in dense regions such as cluster cores. 
This picture can be extended to groups as well, which are found to become efficient in quenching
star formation not earlier than $z\sim$2, resulting in to a blue galaxy fraction declining
down to $z\sim$1 (Gerke et al. 2007).
This accelerated assembly of the more massive systems is at odds with current 
hierarchical models if simply extended also to the baryonic component within DM haloes.

One of the traditional tests for the galaxy formation theories is the colour-magnitude 
relation (CMr) 
and in particular the building-up of the so-called Red Sequence (RS) of early-type galaxies.
The latter is operatively defined as a locus of the CM plane
populated by objects previously selected either by colour or by morphological type.
Clusters are dominated by bright, massive ellipticals 
forming a tight RS  
(Bower, Lucey \& Ellis 1992; Gladders {\it et~al.}\ 1998; Andreon 2003; Hogg {\it et al.}
2004; McIntosh {\it et al.} 2005b), which is classically interpreted as a 
mass--metallicity relation: its slope would then mainly 
driven by metallicity (brighter galaxies are redder because more metal enriched), whereas 
its scatter would mirror the natural spread in galaxy ages (Kodama \& Arimoto 1997).

This mass--metallicity relation holds even at high $z$:
lower mass galaxies 
are metal poorer being still under construction, while more massive ones have already reached
solar metallicity at $z\sim 1$, after having completed the bulk of their star formation 
(Savaglio et al. 2005). 
Claims of either young or coeval stellar populations in cluster ellipticals (see Cole et al. 2000)
were quite ruled out by
findings of a ``anti-hierarchical'' age--mass relation in which more massive galaxies
are older and metal-richer (Nelan et al. 2005): this led to favour a picture where RS proceeds
from more to less massive galaxies, being truncated or depleted in its faint-end at high $z$.

The seemingly universal nature of the RS (see McIntosh et al. 2005a for nearby clusters) 
tends to provide a fair match with
down-sizing evidences over the hierarchical (or bottom-up) merger model (see Chiosi \& Carraro 2002, 
Peebles 2002).
Within the latter, ellipticals have formed as final result of mergers or accretion
processes of smaller objects; in particular by means of `dry-mergers' between gas-free
galaxies having already terminated their activity, thus excluding too recent bursts of 
star formation (see Naab et al. 2006).
 
Semi-analytical models generally tend to accredit a scenario in which
ellipticals keep forming through mergers of late-type disk galaxies, placing themselves on a
RS: here, the RS would be arising from more massive galaxies being also more
metal-rich, because resulting from more
massive merging progenitors (Kauffmann and Charlot 1998, De Lucia et al. 2004a).
Yet, in this view roughly
half of massive early-type galaxies must have assembled from $z<$1, whereas their number is not
reported to decrease out to $z\simeq$0.7-1 (Scarlata et al. 2007). Moreover, this would be
in contradiction with, e.g.,
optical galaxy counts back to $z\sim 3$ (Rudnick et al., 2001), which are consistent with 
large ellipticals being conserved since early epoch, provided that their $B$-band luminosity
were 3 times larger than at $z$=0.

However, De Lucia et al. (2004b) find that a formation framework in which all red galaxies
in clusters evolved passively after a synchronous monolithic collapse at $z\gsim 2-3$,
is inconsistent with observations.
In fact, in a wider concept of hierarchical models, star formation (completed at $z\sim 2-3$ in
higher density systems) may have preceded the actual mass assembly at around $z\sim 1$,
in such a way that star formation occurs ``from top to down'' within DM haloes, 
which are assembled following a ``bottom-up" evolutionary path (see Baugh et al. 1996).
Bower et al. (2006) and Croton et al. (2006) pointed out that the hierarchical CDM model
still provides good match to observational evidences of ``anti-hierarchical'' galaxy
formation,
provided that AGN feedback is taken into account to quench cooling flows in massive haloes.

In this context, the understanding of the building of the RS is a key test for galaxy formation theories 
(see e.g. Cimatti et al. 2006).
Few works have tried to reproduce the observed RS by means of hydrodynamical simulations, so far
(Romeo et al. 2005; Saro et al. 2006), while more numerous are the predictions by semi-analytical methods
(e.g., De Lucia et al. 2004b, Kaviraj et al. 2005); the latter was also the only one to extend the 
analysis to high redshifts, although within an a priori merger paradigm. 
Here we use a suite of TreeSPH simulation as descibed in a series of previous articles (Romeo et al. 2006, hereafter PI;
Romeo et al. 2005, hereafter PII; Sommer-Larsen et al. 2005, hereafter PIII). 
Up to now, such simulations 
have been analysed only at $z=0$, where we addressed, respectively, the intra-cluster medium (PI), 
the cluster galaxies (PII) and the intra-cluster stars (PIII). 

With respect to the previous papers we have now extended the data set to groups of galaxies,
which are by far less studied both observationally and in simulations than clusters:
these groups have already been analysed in D'Onghia et al. 2005 
and Sommer-Larsen 2006, in relation with their possible fossil nature and the intra-group
light content. In this paper we start to exploit our data set in both the
directions of higher redshift and different environments (clusters and groups), in relation
with the same photometric properties as in PII, concentrating for now on the CMr.

In the next section we briefly summarise the simulations and the observational parameter set 
extracted from them. In Section 3 we discuss the appearance and establishment of the RS, with
focus on the assembly of the central galaxies in clusters and groups, and the variation of
its slope with time, along with discussing possible mechanisms able to explain the shape and
evolution of the RS, such as age, metallicity and star formation rate.

\begin{table*}
\caption{Numerical and physical (last three columns, referring to $z$=0) 
characteristics of selected objects: mass of DM/gas/star particles and their respective softening
lenghts; total number of particles and initial redshift of each run; virial mass, radius and
temperature of simulated objects.}
\begin{tabular}{c c c c c c c c c c c c}
\hline
run &	$m_{DM}$ &  $m_{gas}$ & $m_{*}$ & $\epsilon_{DM}$  &  $\epsilon_{gas}$  &  $\epsilon_{*}$  &
$N_{tot}$  &  $z_i$ &  $M_{vir}$  &  $R_{vir}$  &  $kT$ \\
	&  & [$10^7 M_{\odot}/h$] &  &  & [kpc/$h$] &  & & & [$10^{14}~M_{\odot}$] &
	[Mpc] &    [keV] \\
\hline	
C2	& 180 & 25 & 25 & 5.4 & 2.8 & 2.8 & 950000 & 19 & 12.4 & 3 & 6 \\
C1	& 180 & 25 & 25 & 5.4 & 2.8 & 2.8 & 260000 & 19 & 2.8 & 1.7 & 3 \\
C1-HR   & 23 & 3.1 & 3.1 & 2.7 & 1.4 & 1.4 & 2235000 & 39 & 2.8 & 1.7 & 3 \\
12gr.   & 180 & 25/3.1 & 3.1 & 4.8 & 2.5/1.2 & 1.2 & 300000 & 39 & 1 & 1.2 & $\sim$1.5 \\
12gr.-HR & 23 & 3.1/0.78 & 0.78 & 2.4 & 1.2/0.76 & 0.76 & 1500000 & 39 & 1 & 1.2 & $\sim$1.5 \\
\hline
\end{tabular}
\end{table*}

\section{The simulations}
The code used is a significantly improved version of
the TreeSPH code we used previously for galaxy formation simulations 
(Sommer-Larsen, G\"otz \& Portinari 2003). 
Full details on the code and the simulations are given in PI.

The groups and clusters were drawn and re-simulated from a dark matter 
(DM)-only cosmological simulation run with the code FLY
(Antonuccio {\it et al.}, 2003), for a standard flat $\Lambda$ Cold Dark 
Matter cosmological model ($h=0.7$, $\Omega_0=0.3$, 
$\sigma_8=0.9$) with $150 h^{-1}$~Mpc box-length. 
When re-simulating with the hydro-code, baryonic 
particles were ``added" to the original DM ones, which were
split according to a chosen baryon fraction $f_b=0.12$. 

We selected from the cosmological simulation 
one cluster of $T\sim 3$ keV (C1: see PI and PII), one of $T\sim 6$ keV (C2: see PI and PII)
and 12 groups ($T\sim 1.5$ keV), four of which are fossil.
Here it is worth reminding that a fossil group (FG) is defined as a group in which a large elliptical galaxy embedded in a X-ray halo dominates the bright end of the galaxy luminosity function, with the second-brightest group member being at least 2 R-band magnitudes fainter (see D'Onghia et al., 2005).
TreeSPH re-simulations of cluster C1 were run with different super-wind (SW)
prescriptions, IMFs, with or without thermal conduction, with or without 
pre-heating (see PI). 

In order to allow direct comparison of the results
pertaining to the different targets, in this paper only the ``standard" model SW-AY
will be exposed for all of them. This scheme makes use of an Arimoto--Yoshii (AY) IMF which 
is top-heavier with respect to a classical Salpeter one, along with a feedback
prescription in which 70\% of the energy feedback from supernovae type II goes into driving
galactic super-winds (SW). 
As for the IMF, Saro et al. (2006) find that a Salpeter IMF successfully reproduces
the amplitude of the {\it V-K} vs.{\it  V} relation at the bright end ($V\lsim -20$), producing
though too blue faint galaxies; while a top-heavy IMF would produce too metal-rich galaxies,
raising the overall level of the relation.
Yet, Portinari et al. (2004) warned that a Salpeter
IMF is unable to account for the level of metal enrichment in clusters of galaxies, which would
instead be attained by a `non-standard' IMF, through a lower locked-up metal fraction in stars.

However, the SW-AY is not a neutral choice, since by combining the
results for the gas in PI and for the galaxies in PII, it turns out that a unique
model able to describe at the same time and in a satisfactory way both the ICM and the galaxy properties,
does not apply: while choosing a strong feedback {\it and} a top-heavy IMF brings, by 
efficiently counterbalancing cooling, 
to successfully reproduce the correct X-ray luminosity - temperature relation, entropy
profiles and cold gas fraction of the clusters, on the
other hand all this occurs at expenses of insufficient star accumulation,
causing a deficiency of $M^*$+2 galaxies. 

%

A drawback common to most known hydrodynamical simulations, 
where AGN feedback is not properly assumed,
is to produce an apparently bluer CD (or brightest central galaxy, BCG), as a result of 
an accelerate late stellar birth-rate even after the epoch of quiescent star formation:
after the cluster has recovered from the last major merging events,
a steady cooling flow is established at the centre of
the cluster, being only partially attenuated by the strong stellar feedback;
it gets turned in to young star populations,
significatively contributing to the total luminosity.
As in PI and PII, a correction is then applied, by
removing the star particles formed within the innermost 10~kpc of the BCG since $z_{corr}$=2,
epoch when the bulk of cluster and group  galaxies have already formed.
%

Finally, in order to increase the resolution of the stellar component in 
group simulations, we chose that
each star-forming SPH particles of the initial mass was 
gradually turned into a total of 8 star-particles. Thus SPH particles, which
have been formed by recycling of star-particles, will have an eighth
of the original SPH particle mass.

To test for numerical resolution effects, the cluster C1 was also run with eight 
times higher mass and two times higher force resolution.
As a result of the increased resolution, in this run the LF will extend to fainter
magnitudes, since substructure is here more resolved in form of higher number of 
proto-galaxies (see PII). In the following, then, all the plots pertaining to
the two clusters will reach a completeness limit which is higher with respect
to the analogous ones referring to the groups, where the dwarf population is not
resolved as well as there.
 
The particle masses and softening lenghts of all simulations
are resumed in Table 1.

\subsection{The luminosity and colours of identified galaxies}

The individual galaxies are identified in the simulations by means of the procedure 
detailed in PI and PII. They are assigned 
a luminosity as the sum of the luminosities 
of its constituent star particles, in the broad bands UBVRIJHK (Johnson/Cousins filter).
Each star particle represents 
a Single Stellar Population (SSP) of total mass corresponding to $m_*$ (see Table 1),
with individual stellar masses distributed
according to an AY IMF; each of these SSPs remains characterized by its age and 
metallicity over a redshift range, from which
luminosities are computed by mass-weighted integration of the Padova 
isochrones (Girardi {\it et al.}, 2002).

In particular, each star particle of mass $m_*$ corresponds to an
initial SSP mass $\frac{m_*}{1-E(t)}$, where $E$ is the fraction of the star particles  
which transforms back into gas particles after a time $t$. Therefore, for the computation of the
luminosity each star particle is assigned such a corresponding ``initial SSP mass'', rather than 
its actual present mass $m_*$ (see PII).

The step of defining the total magnitude of the simulated galaxies is the most critical when trying
to provide galaxy properties that might compare to observational quantities. Even though, by definition, 
the total luminosity is given by the sum of the luminosities of all the stellar particles belonging to a 
galaxy, there remains an arbitrariness on the bound criterion definition. Assuming that our procedure is 
correct, there are a few cases where we had to pay some attention to the quantities derived by the bound 
particles --first of all BCG galaxies.
These systems show generally a very extended stellar halo (typical of CD galaxies: 
Bernstein et al. 1995),
somehow larger than typical fixed aperture magnitude estimates used for the observed systems. Aperture 
correction is typically applied to the observed systems in order to compensate the missing flux and
have an unbiased total luminosity estimate, but it does not necessary guarantee unbiased colours.

For the simulated BCGs we have figured out from the internal colour
profiles that the larger is the distance from the galaxy centre the bluer is its total colour 
within.
We decided to use fractional apertures in order 1) to cut the BCG envelope (not 
properly belonging to the galaxy) out and 2) to estimate colours in a way consistent with observations. 
The use of fractional apertures has the caveat of leading to a flattening of the RS and an increase 
of its scatter, as Scodeggio (2001) showed for Coma.
This effect becomes significant from $M_V\lsim$-21.5 up, so that we chose to apply it
to the BCG only. Instead, magnitudes are always computed on the total particles once got rid of 
the diffuse envelope.

For all the other galaxies we have found that the criterion of the bound particles is accurate both for 
total luminosity and colours. For the very faint part of the luminosity (and mass) function, we put a 
minimal threshold $N_{\rm par}=15$ of bound particles in order to define a galaxy. This has been conservatively 
chosen in order not to loose too many simulated galaxy candidates and, complementarily, not to
over-predict intracluster/group star particles.
Once compiled the galaxy catalogue, we can always discard the objects which are beyond the 
luminosity (and mass) completeness limit from the analysis, when necessary. 

\subsection{Definition of the galaxy samples} \label{samples}

All the obtained galaxies were considered as belonging to their pertaining environment. 
A temptative classification of environments can be given by the radial galaxy density, so that
we separate all cluster galaxies in two classes corresponding to regions inside and outside
a threshold radius of
$1/3R_{vir}$. All the 12 groups were considered in the same class of richness
and were only classified by their normal or fossil nature. As a result, we have four
classes of galaxies, which we will refer to as IN (2 cluster cores), OUT (2 cluster outskirts),
NG (8 normal groups) and FG (4 fossil groups).

The inspection of the luminosity function (not discussed in this paper) shows that, relatively
to $V$-band, galaxies are complete at $M_V=-18,-19,-20$ for $z$=0, 0.5, 1 respectively and stays complete 
at about $M_V=-20$ out to $z$=2, independently on the environment (cluster or groups). These values 
correspond approximatively to galaxy stellar masses of 
$\sim 10^{10} M_\odot$ for C2, $3\times10^{9} M_\odot$ for C1 (higher resolution) and
of $4\times10^{9} M_\odot$ for the 12 groups (which have higher stellar particle mass resolution -see above), 
at all redshifts. Thus, the simulated samples are always complete at sub-$L^*$ luminosities, and mass 
resolution is enough to explore also the luminosity and mass range down to the dwarf galaxy regime, even 
though we will not be complete here.

A final consideration is needed about the definition of the early-type (ET) subsample. 
This is the simulated 
galaxy sample to be used in order to perform some basic analysis like the measurment of the 
slope, scatter and zero-point 
of the RS to be compared with the observations.
The simulation set-up does not allow to deal with galaxy morphology (in this sense we are in the same 
situation of many ground-base large scale structure surveys). For this reason we decided to adopt a 
selection based on colour -- internal velocity dispersion, which can be consistent with ET objects. 
ET systems were considered as those galaxies brighter than the completeness magnitude at each 
redshift, lying within $\sim 2\sigma $ around the median colour and having internal velocity dispersion 
larger than 100 kms$^{-1}$. Once defined the ET sample and identified a fiducial RS, a linear 
best-fit is computed in order to characterise the evolution of the sequence with time, as we will discuss later on.

\begin{figure*}
\centering
\includegraphics[width=8.5cm]{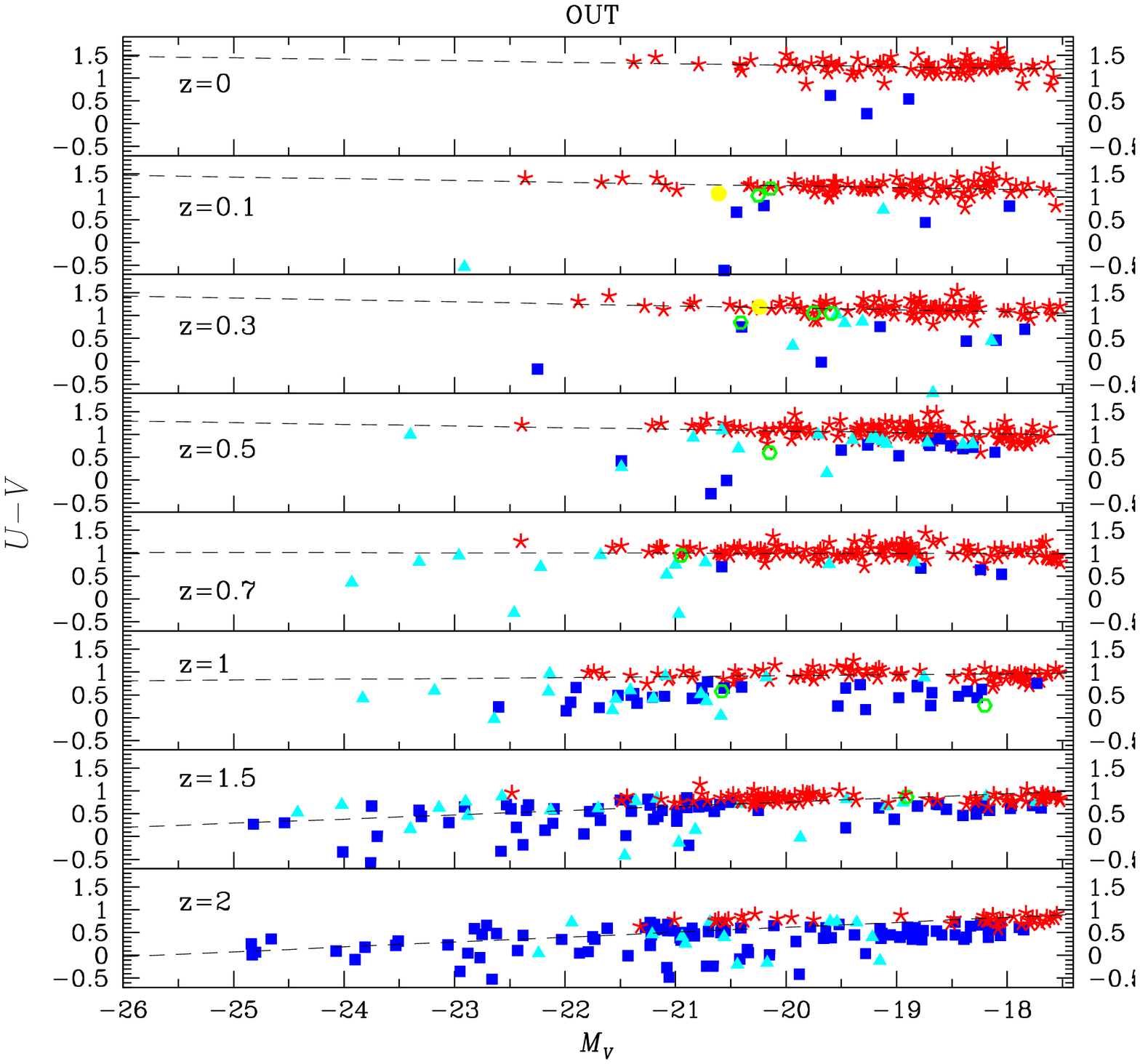}
\hfill
\includegraphics[width=8.5cm]{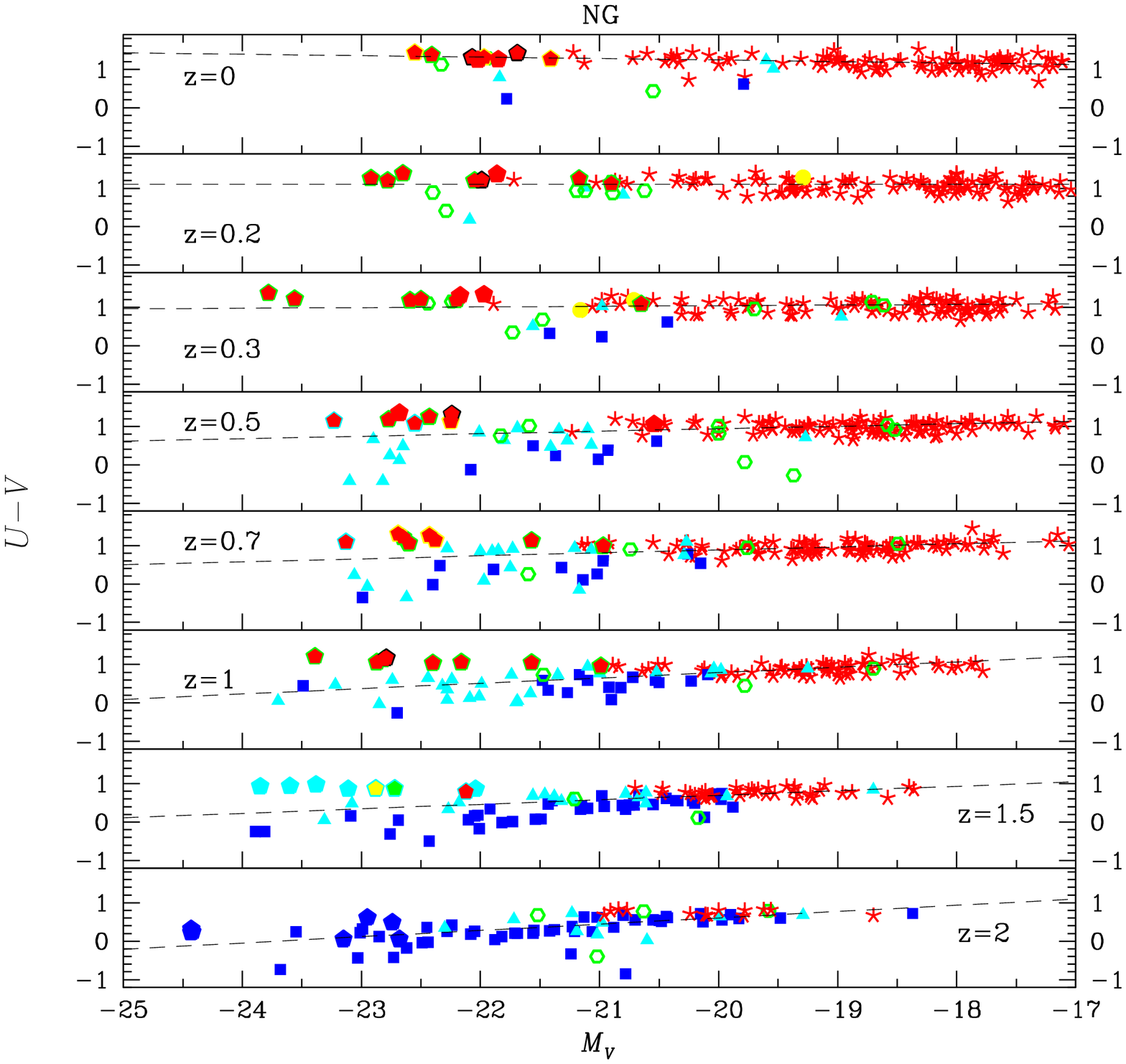}
\hfill
\includegraphics[width=8.5cm]{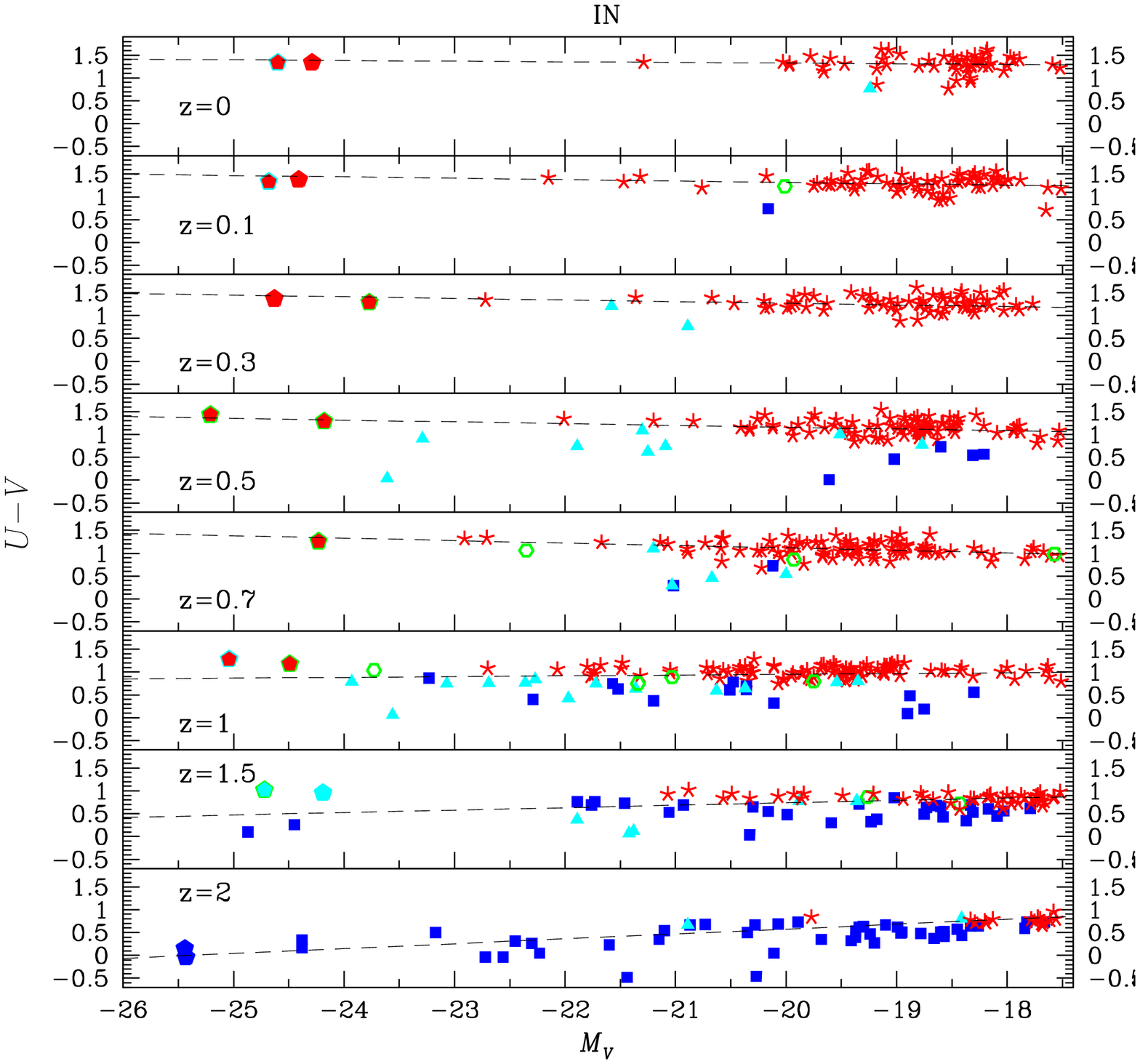}
\hfill
\includegraphics[width=8.5cm]{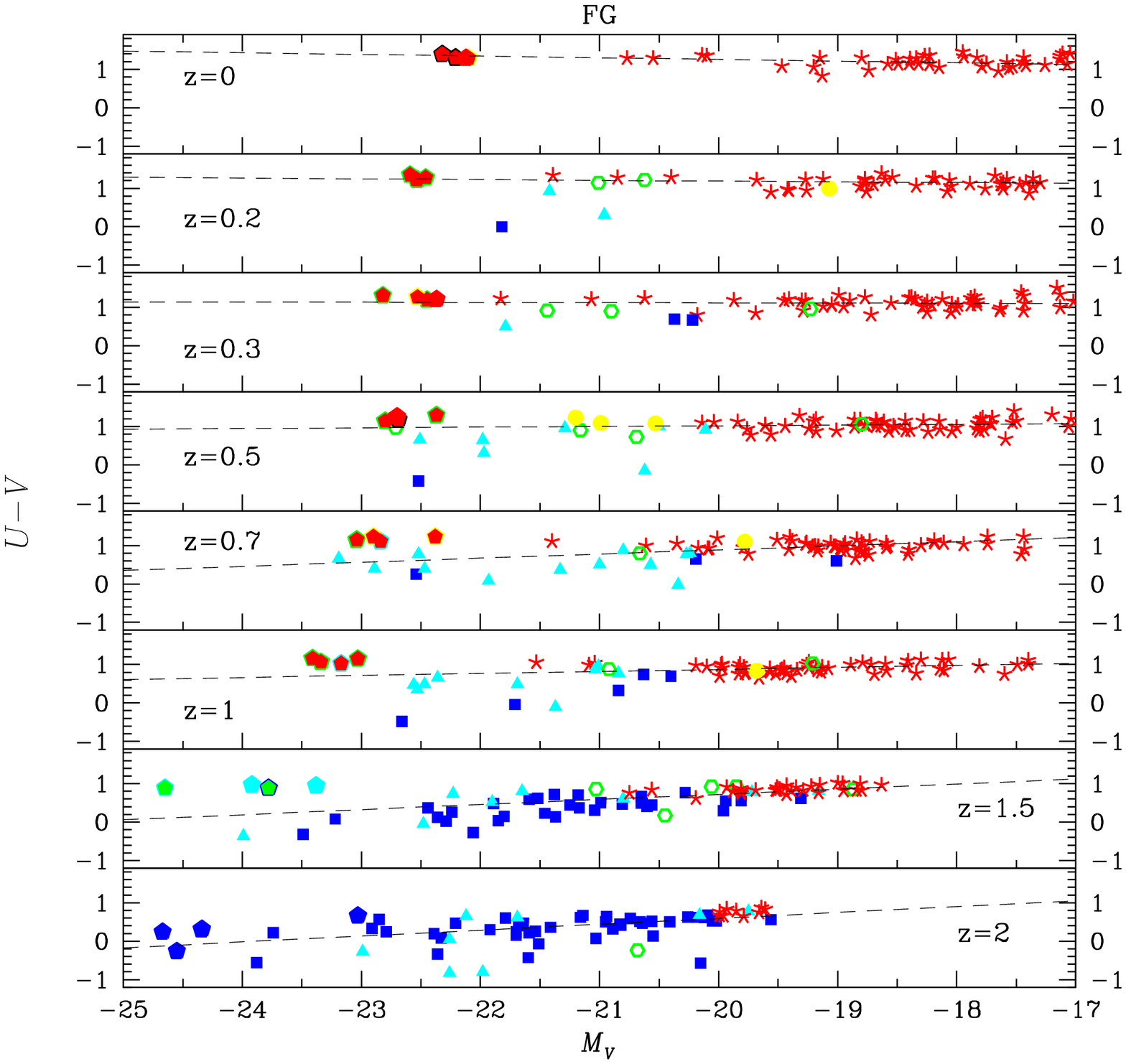}
  \caption{{\it U-V} vs. {\it V} CMr for cluster galaxies outside (upper left panel)
  and inside (lower left) the core, and in normal (upper right) and
  fossil groups (lower right) at different
  redshifts. Galaxies are classified by their specific star-formation rate (SSFR): 
  $<$-4 (red stars), -4-- -3 (black), -3-- -2 (yellow circles), -2-- -1 (green open circles), 
  -1-- -0.3 (cyan triangles), $>$-0.3 (blue squares) in units of log$(M_{\odot} Gyr^{-1}/M_*)$.
 Aperture-corrected BCGs are shown as filled pentagons, while their whole colour (including
the cutted envelope) is shown as contour. } 
\label{uv}                  
\end{figure*}

\section{Results}


The {\it U-V} vs. {\it V} colour-magnitude diagram of simulated galaxies is shown for the cores 
and outskirts of the clusters C1 and C2 and for the 12 groups in Fig.\ref{uv}. 
Galaxies are here coloured according to
their specific star-formation rate (SSFR=SFR$/M_*$) over the last previous Gyr. 
Data points are accompanied by the linear fits to 
the ET subsamples, as defined in the previous section. 

We cumulated galaxies belonging to same 
environment in order to derive average galaxy properties on significant statistical samples. 
The groups (divided in normal and fossil ones) are fairly homogeneous since they have been 
selected within a 
narrow range of virial mass and temperature while the two clusters were significantly different 
with respect to
the same virial quantities (see Table 1). This will allow us to test the CMr properties of 
galaxies in a wide 
range of global cluster and group properties (one order of magnitude in mass, 3 times in 
radius and a factor of 4 in temperature).

We have decided to adopt a standard diagnostics, i.e. {\it U-V} colours 
vs. {\it V} magnitude, because the {\it U-V} colour allows to sample the 4000\AA\ break
of ET galaxies and is therefore particularly sensitive to age and metallicity variations 
of the stellar populations in
galaxies, and for consistency with rest-frame bands used in previous works in literature 
(Sandage \& Visvanathan 1978, Schweizer \& Seitzer 1992, Bower, Lucey \& Ellis 1992, 
Terlevich et al. 2001, Bell et al. 2004, De Lucia et al. 2004a, McIntosh et al. 2005a, Tanaka et al. 2005, 
Wolf, Gray \& Meisenheimer 2005: see subsection \ref{obs}). 

\begin{figure*}
\centering
\includegraphics[width=8.5cm]{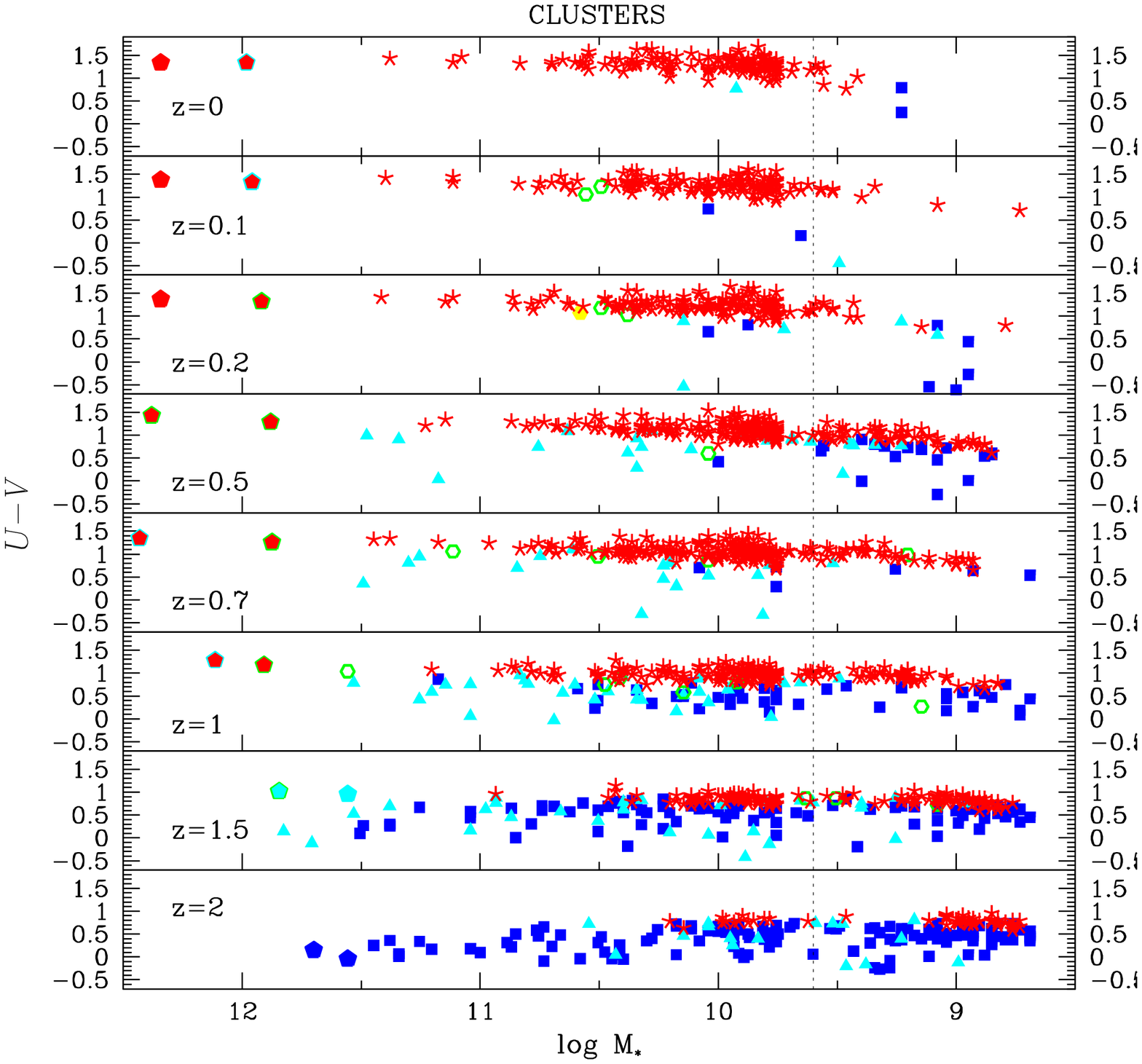}
\hfill
\includegraphics[width=8.5cm]{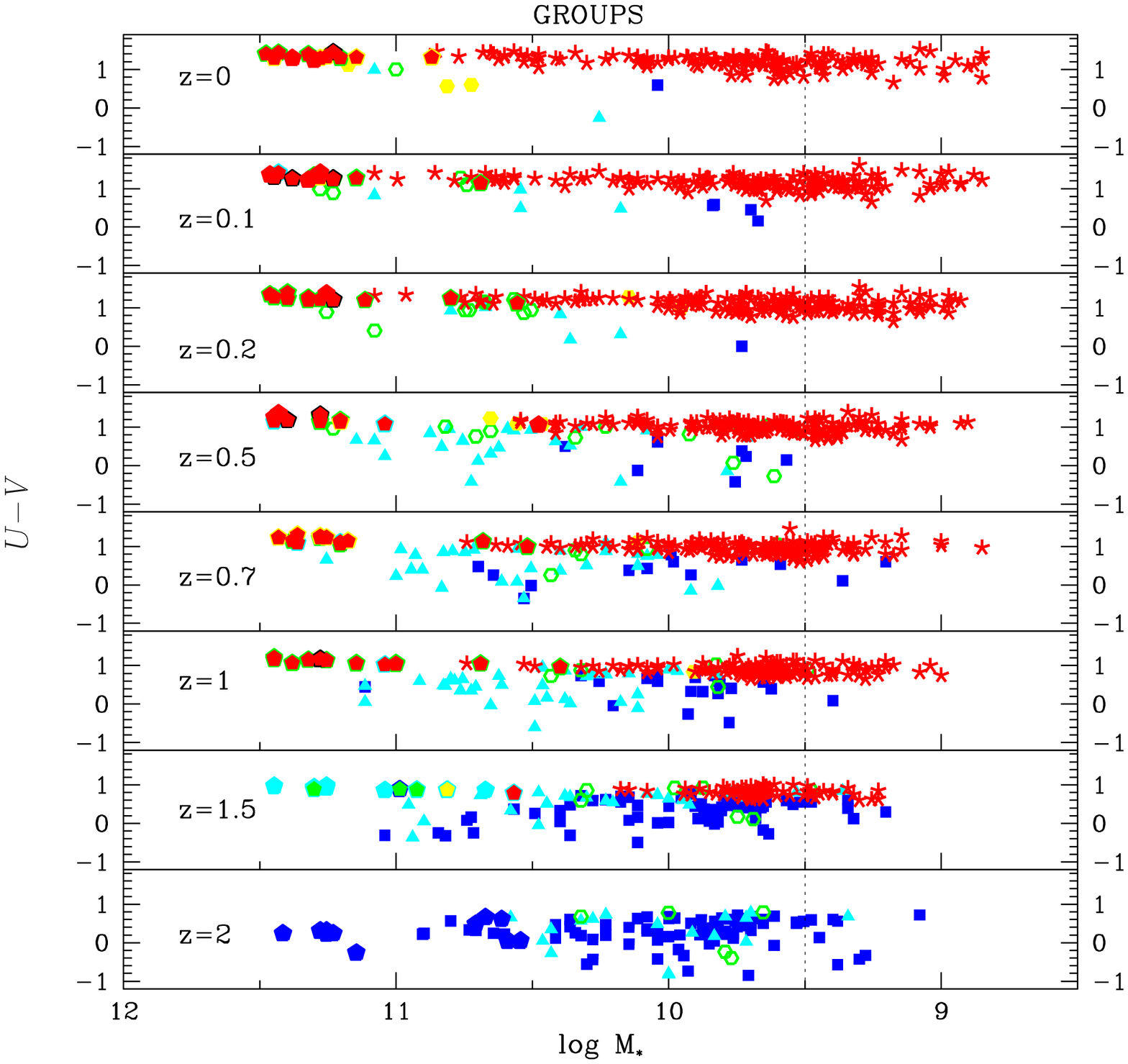}
\hfill
  \caption{{\it U-V} vs. stellar mass for the same objects as in Fig. 1, now collected in two classes: 
clusters (left) and groups (right), at different redshifts. Symbols are as in Fig. \ref{uv}. 
Completeness limits (dotted line) are calculated for cluster C1 and for the groups separately (see Sec. \ref{samples}). } 
\label{uvms}                  
\end{figure*}

\subsection{Building of the Red Sequence in clusters and groups} \label{buildup}


Over time the RS (as marked by the linear fit to the ET sample) seems to have a similar evolution 
regardless the environment: it seems to stay in 
place since $z=1$, epoch at which also the BCGs move to the red region of the 
CMr already occupied by those less massive galaxies which are already red even at earlier epochs. 

All galaxies in the CMr behave according to a main direction: blue star forming galaxies 
get redder as soon as 
they start quenching their star formation and progressively move toward a sequence of ``dead'' 
or quiescent galaxies,
which we can hereafter refer to as ``quiet or dead sequence'' (DS), 
having no detectable star formation. This sequence starts to arise from the faint end since 
early epochs ($z\sim1.5$) in the same way in all the environments. At that epoch bright galaxies in clusters 
(for instance those within 3 magnitude from the BCGs, $M_V<M^{\rm BCG}_V+3$) show high star formation 
activity, which is reduced later on much quicker than for fainter systems, where significant star formation 
holds until $z=0$ in a few cases. The same behaviour does not seem to be present in groups, where brighter 
systems keep forming stars in the same way as the faint ones till lower redshifts. 

The main features of the RS in the colour-magnitude plane as they are found in Fig.\ref{uv} are 
essentially the following:
\begin{itemize}
\item{All star-forming galaxies lie below the RS fits since $z$=1.}
\item{The RS gets progressively completed from the faint end, where those galaxies gather that
have first shut their star formation off.}
\end{itemize}
The main difference in building the RS between normal and fossil groups 
stems from the different timescales of star formation activity and in the mass assembly of the central 
galaxy: whereas in the latters the RS is to get completed before $z$=1, it is still under construction 
at $z$=0.5 in normal groups. 
Indeed, if bounding ourselves to the normal groups, here
the CMr seems to evolve with a slowed down pace with respect to clusters: galaxies in groups are 
still in the phase of moving towards
the RS at epochs when the latter is already in place for galaxies in clusters; under this respect,
the most marked difference is found between normal groups and cluster cores, while fossil groups
rather behave like the latters.

In Fig. \ref{uvms} we can have the perspection of such process in terms of the total galaxy stellar 
masses. Roughly splitting the sample by drawing a line at $\log M_*=10.3$, one finds that in  
clusters the largest contribution to the total star formation is given by
low massive galaxies, while in groups there 
remain active galaxies until $z=0$ quite uniformly distributed at all masses.
 
As a consequence of that, there seems to be some evidence that the a ``downsizing'' mechanism, by which 
the early star formation activity of the brightest systems is interpreted as antihierarchical, is at
work in the galaxy stellar component evolution within clusters' dark matter halos, whilst less for groups, particularly the normal ones.
This could be due to a different galaxy activity between cluster and group environments. For 
instance, later than $z$=0.5 star formation is about over in the central cluster regions except for some minor 
activity around the BCGs or galaxies falling into the central regions from the outskirts. 
On the lower density environment side, there are proportionally more active galaxies in cluster 
peripheries and normal groups. 
On their own, fossil groups witness a more drastic reduction of the activity already at $z=0.3$. 
However, it is remarkable 
that active galaxies in groups occupy the bright part of the CMr as well as the faint ones. 

Our simulations can not allow a detailed treatment of the physical processes that concur to a
kind of density dependent inhibition, 
nor this is one of the issues of the present paper. We have however checked that one of the major 
drivers of the earlier quenching of 
the star formation is the fact that core cluster galaxies show smaller or even null gas fractions, 
which suggests that they 
have been gas stripped as a consequence either of ram pressure or galaxy-galaxy interactions.


\subsection{The r\^ole of star formation through environments} \label{env}

\begin{figure}
\centering
\includegraphics[width=8cm]{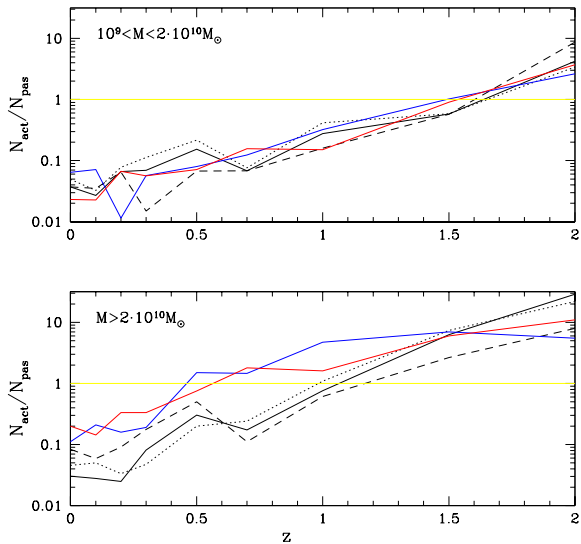}
  \caption{Ratio of star-forming over non star-forming galaxies in 
  cluster cores (dashed lines), cluster outskirts (dotted),
  overall clusters (solid black), normal (blue) and fossil (red) groups, as a function of
  redshift, for objects less (top) and more (bottom) massive than $2\cdot 10^{10}M_{\odot}$.}
\label{ngal}		  
\end{figure}

The presence of a transition epoch between star-forming to passively evolving systems is 
made clearer by complementing the 
CMr with the variation of the stellar mass in galaxies shown in Fig.\ref{ngal}. 
Here galaxies are divided in star-forming and 
non star-forming by the threshold SSFR$=1 M_{\odot} yr^{-1}/10^{10}M_{\odot}$
and classified as more (lower panel) and less (upper) massive than $2\times 10^{10}M_{\odot}$.

The overall trend points towards a diminution of the active galaxy population over the 
last 10 Gyr, with no significant variations among the less massive objects due to environment. 
Conversely, the quiescent population globally experiences a steadily increasing growth 
in number, especially in the less massive galaxies, where it roughly doubles in all environments
at an epoch even earlier than $z$=1.

When considering the more massive galaxies, the active-to-passive ratio 
in clusters decreases at a constant rate between $z$=2 and 1, which is steeper than in groups.
Indeed in normal groups the ratio is almost flat until $z$=1, after which it begins to decrease.

At $z$=0.7 the active-to-passive ratio in clusters undergoes an inversion, which lasts until $z$=0.5:
this is due to the increase of the number of less massive active galaxies in the outskirts, mirrored by
an analogous uprise of massive active galaxies in the cores.
This behaviour is consistent with the occurrence of some merging phenomena involving satellite
systems which induce star formation in the BCG outskirts.

Later on, a peak occurs around $z$=0.5 of the high mass active galaxy number in the
cluster cores and of low mass active ones in the outskirts.
This is a consequence of a high number of infall/pass-by events during which
galaxies can eventually get activated because of interaction with the ICM (see S\'anchez et al., 2007).
A steep decline of massive active galaxy number from $z$=0.5 on is also outlined in normal
groups, while galaxies in fossil groups follow a smoother evolution towards reddening.

In particular, considering the ratio between more and less massive quiescent galaxies, we calculated 
that it maintains almost constant ($\simeq$0.2) from $z\simeq$0.7 to 0, not reproducing thus that deficit
of faint red galaxies found in cluster cores at $z$=0.8 by De Lucia et al. (2007) and Stott et al. (2007), 
but questioned by Andreon (2006 and 2008): more light can be shed on this by studying the evolution
of the ``dead'' galaxy LF in terms of the dwarf-to-giant ratio.

A strong estimator of star formation variancy through environments is given by the epoch
at which the inversion of the ratio between active to passive regimes occurs: among more
massive systems, it is higher for cluster environment, 
where this happens at $z=1.2$ in the 
central regions and $z=1$ in the outskirts, than for fossil groups ($z\sim0.55$) and normal 
groups ($z\sim0.45$). 
This difference is even more marked for the redshift when the active galaxies become less 
than 50\% of the passive systems --which are 
$z=0.8,~0.4$ and $0.3$ for cluster, fossil and normal groups respectively. 
This supports the validity of an environmental sequence of
star formation activity, as overviewed in Table 2. 
Moreover, such a sequence marks also a difference in the processes driving the 
evolution of the fossil against normal 
groups at later epochs, where the gap arises in the formers between the BCGs and 
the second ranked luminous galaxies.
On the contrary, systems with 
$M < 2\times 10^{10}M_{\odot}$ all invert the active-to-passive ratio at $z\simeq 1.5$, 
regardless of the environment.

\begin{table}
\begin{tabular}{l c c c c}
\hline
 		& $z_{act}$ &  $z_{\frac{1}{2}act}$ & $z_{\alpha =0}$ &$z_{ds}$ \\
\hline	
cl.core (IN)	& 1.2 & 0.85 & 0.9 & 0.9\\
clusters (whole)	& 1.05 & 0.8 & 0.8 & 0.75 \\
cl. OUT   	& 1 & 0.8 & 0.7 & 0.5 \\
FG   		& 0.55 & 0.4 & 0.4 & 0.35 \\
NG 		& 0.45 & 0.3 & 0.2 & 0.25 \\
\hline
\end{tabular}

\caption{Environmental dependency of the transition epochs 
at which the number of active galaxies equals the number of passive ones ($z_{act}$),
and at which is half of it ($z_{\frac{1}{2}act}$), as of Fig. 3, 
for galaxies having $M>2\cdot 10^{10}$,
compared with 
the epoch at which the slope $\alpha$ of the RS gets null ($z_{\alpha =0}$) and at
which it equals that of the DS ($z_{ds}$), as of Fig. 5.}
\end{table}

The evolution of the ratio between the 
active and passive systems is closely specular to a similar evolution of the RS 
in terms of slope and scatter, regardless the environment at least until 
z=1, as we will detail more quantitatively in Sec. \ref{slope}.
Only after this epoch the galaxy activity clearly differs between clusters 
and groups and the environment starts to play a significant role.

The overall picture aforedrawn is substantially in agreement with recent
findings by Arnouts et al. (2007), who fix at around $z$=1.2 the transition epoch 
between the active to passive field galaxy populations over a wide stellar mass range 
($K_{ABS}>-23$). 
At that time the active population was already in place with no other evolutionary
mechanisms than the ageing of its SSPs, while the quiescent sequence shows a gradual 
increase by doubling its stellar mass, at the same rate as new stars are formed. 
This period in which active galaxies turn off the bulk of their star formation, finally 
coincides with the epoch
of establishment of the RS, after which further stellar mass assembly only involves 
isolated merger episodes mostly on to the BCGs. Their $z=1.2$ transition epoch stays 
comfortably in between our high limit from low 
mass systems ($z=1.5$) and the low limit from high mass systems ($z\sim 1$).

Also promising is the consistency of our results with the evidence coming from distant 
cluster surveys 
pointing out that the red passive population has begun to shift towards the RS when their 
activity came to an end at around $z\lsim 0.8$, possibly as a result of environmental 
effects (Smail et al. 1998, Poggianti et al. 1999). 
As seen in Fig. 1, our simulated clusters (both in central and outer regions) show in 
the very interval between $z=1$ and 0.7 a dramatic decrease of the star-forming systems 
in the bright part of 
the CMr, which goes along with the reduction by about half of the active population 
in the same period as shown in Fig. 3.
A relevant part of the low massive systems merge on to the CD,
but many small systems
get destroyed when passing through the cluster centre and end up with feeding
the intra-cluster light (which indeed we calculated to grow significantly in the same redshift range).


The evolution of the colour distribution of the quiescent and active galaxy population 
will be discussed further in section \ref{coldis}.


\begin{figure}
\centering
\includegraphics[width=9cm]{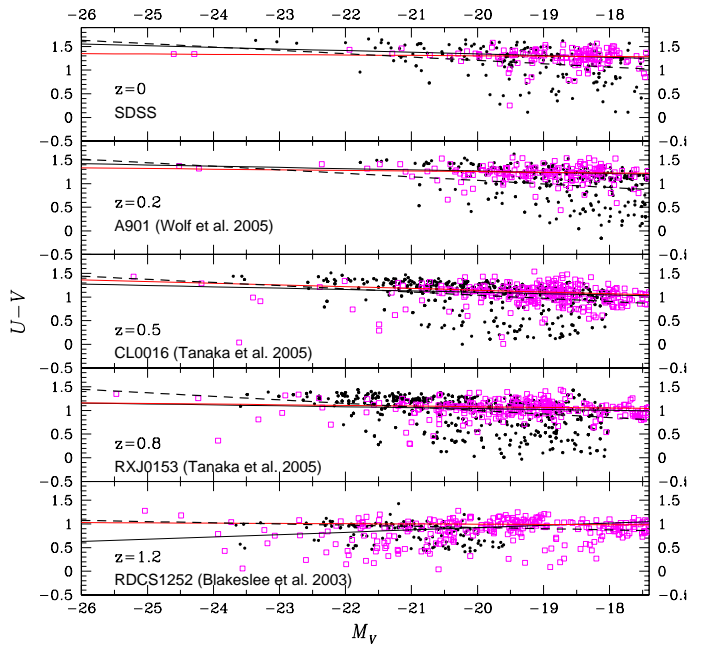}
\hfill
  \caption{Rest-frame {\it U-V} vs. {\it V} CMr for galaxies in the two clusters (purple open squares): black solid lines
  are fit to the Red Sequence samples, red lines are fit to all non star-forming objects (DS); dashed
  lines are data from observed clusters at each redshift (converted to rest-frame colours), randomly sampled
  to the same number of simulated objects (black points). SDSS data are taken from Tanaka et al. (2005).}
\label{cmobs}
\end{figure}

\subsection{Comparison with observations} \label{obs}

In Fig.\ref{cmobs} we compare the cumulated CMr data of the two simulated clusters at 
redshifts 0, 0.2, 0.5, 0.8, 1.2, with observed samples taken at approximately the same redshifts. 
The observational samples have been 
chosen in order to match the $U-V$ rest-frame colours and magnitudes of the simulated sample 
so that no correction had
to be applied for the comparison.
The best-fits to the observed RS provided from the authors (dashed lines) are
compared with the linear fits to our RS (continuous black line) and DS (continuous red line):
the RS and DS objects constitute two entirely indipendent subsamples, where the former has been 
computed emulating the operative definitions used in the observations (see Sec. \ref{samples}).
Here we recall 
that the fit to the RS is subject to a degree of arbitrariness from author to author due to the 
colour-cut selections and limiting magnitude adopted for the fit (here we adopt the same selection as in 
Fig. \ref{uv}), while our fit to the DS is unrespective of the colour and magnitude (mass) pre-selection, since it 
is by definition applied to the galaxies with a SFR below the limit fixed by SSP resolution.
 

Overall the simulated CMr data seem in fair agreement with observations, not only with respect to their zero-point 
(simulations have on average the same colours of the real galaxies at every magnitudes), but also in their  
global trends: the best-fits to the data generally follow the simulations within the errors.  
However there seem to be some systematic effect: the best-fits lines to the RS of the observed samples deviate 
from the simulated ones at high-$z$ and becomes increasingly coincident with the simulations at low-$z$. 
This can be 
explained with the fact that the observed samples become poorer and poorer than the simulated samples 
going back in 
time, where galaxy surveys tend to favour the brightest members, so that the slopes derived from 
the observed samples 
could be biased toward the more luminous and red systems. 

On the other hand, the proposed DS sample shows a fairly constant slope over times, 
being the reddening of the 
galaxies due to no other processes than passive evolution.
The gap between the two simulation fits (RS and DS) is found to decrease with time, 
as the RS approaches 
the quiescent galaxies line at the latter's quasi-constant slope (see Sec. \ref{slope}): the two end up
coinciding already at $z$=0.8 where the dead galaxies are evidently dominating the RS. 
All in all, the convergence of the RS to the DS as defined in the simulation is a 
further indication that the DS is the sequence which the RS converge to, after having 
quenched the star formation 
activity.

The global trends could be affected by the choice of the IMF and feedback,
as can be inferred (although at lower resolution) from Figs. 4 and 10 of 
Paper II. In case of Sal/SW the luminosity
distribution would get stretched toward higher magnitudes, while colours stay
almost unchanged for $M_B < -19$ (which mainly embraces the range of our RS).
In case of Salpeter IMF with weak feedback, again the net effect is a
brighter luminosity distribution, while colours are similar to those of AY/SW.
According with these indications from $z=0$ we expect that a different
IMF + feedback recipe would alleviate the known lack of brighter objects,
yet leaving the galaxies on the same RS.

The agreement with observational trends, despite lacking physics such as AGN feedback, 
is encouraging in the direction of
well reproducing the colour properties of galaxies out to $z=2$. 
This means that probably AGNs play a 
major role only in high mass systems (of the order of our BCGs down to $M_V^{BCG}+2$). 
Probably the presence of AGN 
activity would have some impact also in very later epochs $z<0.5$, 
where there remain some bright systems ($M_V=-21$ 
to $-23$, in particular for the SDSS sample) redder than predicted, somehow 
pointing towards either 
an earlier and more efficient star formation inhibition or a more efficient 
metal enrichment, as AGN would provide.
This is because the presence of AGN is also expected to improve the metal 
recycling, so that to possibly
solve the small offset in colours shown by simulated galaxies between
$M_V$=-22 and -21 during the period z=0.8 to 0.
These results make us confident about the reliability of the colour estimates of 
galaxies and also the ET subsample selection which will be considered hereafter 
for picturing out the evolution of the RS.

\begin{figure}
\centering
\includegraphics[width=8.5cm]{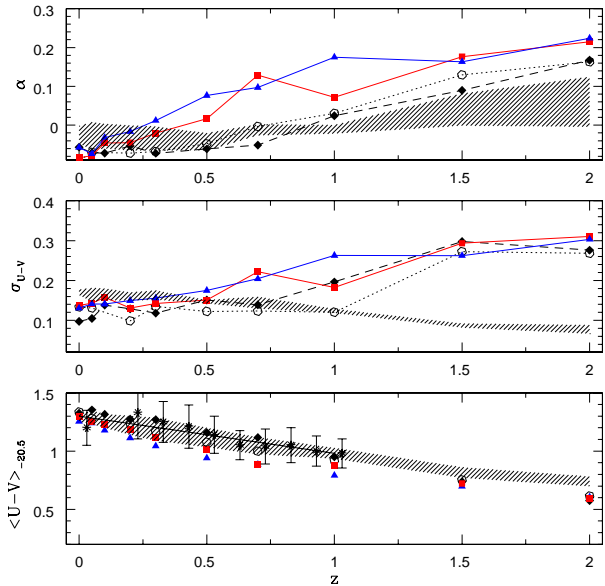}
  \caption{Slope of the linear fit to the {\it U-V vs. V} CMr (upper),
  scatter of the CMr (middle) and its intercept at $M_V=-20.5$ (lower) in the cluster cores 
  (black diamonds), cluster outskirts (open diamonds), normal groups
  (blue triangles) and fossil groups (red squares).
  Data points are compared with the inactive galaxies of the DS (shaded region: all environments,
  all masses). Zero-points are also compared to data from Bell et al., 2004 (asterisks with errorbars). }
\label{slosca}		  
\end{figure}

\subsection{Slope, scatter and zero-point of the CMr}\label{slope}

The characteristic quantities defining the RS are its slope, scatter and zero-point.
The slope of the RS is predicted by hierarchical models to vary with redshift:
Kauffmann \& Charlot (1998) find a slow change leading to a flattening at $z\simeq 1.5$
for clusters ET galaxies formed by merging of  smaller objects.
On the other hand, Kodama \& Arimoto (1997), considering a stellar generation born at high $z$
and purely passively evolving, predict that the CMr changes very little until quite close to that epoch
of star formation.
In their modelling of RS driven by metallicity, based on a monolithic scenario,
Kodama et al. (1998) showed that
the sequence is expected to evolve with a quasi-constant slope,
if taken at rest-frame band in clusters, up to $z$=1.27 (as confirmed by Lidman et al., 2004): 
this is consistent
with a passively evolving population of old ellipticals, with little differential
evolution as a function of galaxy luminosity.
In this picture the differences along
the sequence would be driven by mean stellar metallicities. 

The zero-point together with its scatter can provide hints over the formation epoch of
the galaxies. On one hand, the evolution of the zero-point with redshift gives a measure of how 
the overall 
reddening of the RS is consistent with passive evolution of galaxies. On the other hand, 
the scatter of the RS is a 
measure of the duration of the formation process of the galaxies building-up the RS itself: 
as the stellar populations making up galaxies,
not all formed at the same epoch, get older, the scatter in their colours is expected to diminuish. 
In the local Universe the scatter in the early-type colours is observed to be small ($\lsim$ 0.04 mag 
in clusters: Bower et al. 1992, Terlevich et al. 2001; $\sim$0.1 mag in fields), giving a clue towards a 
not large age spread of this population. 
However, as seen in the previous paragraph, the colour scatter of galaxies is quite affected 
by the star formation activity as far as the RS becomes tighter with time, as a consequence of the 
quenching off of the star formation in galaxies in nearly all the environment. 

Having the opportunity of tracking the colour evolution, one is able to determine when the scatter becomes large, 
and/or at which $z$ the slope of the RS changes: this point where the RS breaks down represents the epoch 
of the major galaxy formation events (i.e. significant star-bursts).
In Fig.\ref{slosca} we show the evolution of the slope, scatter and zero-point of the CMr, for 
galaxies belonging to the 
two previously defined RS (populated by the ET selected systems) and DS (made up of those
galaxies that have almost turned-off their star formation activity).
%

There seems to be a marked difference between the evolution of the RS slope in cluster and groups in top 
panel of Fig. \ref{slosca}. Cluster galaxies slowly change their slope back to $z$=0.7 (even later in the 
outskirts) and then they steeply change the rate of the slope increase back in time. On the other hand, 
groups have a slope which is slowly decreasing from $z$=2 to $z$=1 and quickly decreasing to the typical 
cluster values only around $z$=0.2 at the epoch where the cluster RS already stays stably in place. 
This behaviour is indicative of an earlier formation and faster evolution of the cluster galaxies, 
with respect to groups. There seems to be also a sequence of the trends with the environment since we can 
distinctively remark that the epochs where slopes cross the $\alpha$=0 value fall earlier 
and earlier from cluster centers ($z$=0.9) to cluster outskirts ($z$=0.7) and from fossil ($z$=0.4) 
and normal ($z$=0.2) groups, strikingly coincidents to the ephocs of the active-to-passive inversion 
discussed in section \ref{env} (see Table 2).   

This interpretation is well motivated when considering the evolution of the DS slope, shown as shaded region in 
the same figure panel. The DS stays almost constant over all environments, particularly 
since $z\sim1$ with a slope $\alpha_{\rm DS}$ ranging between -0.06 and 0 and a median value 
$\alpha_{\rm DS}=-0.02$. The ET systems start to behave as dead galaxies progressively later in 
different environments, being cluster core galaxies the first ones to get into the DS regions (at $z\simeq$0.9) 
and the normal groups the latest ones (at $z|simeq$0.25).   


The scatter of the RS is mainly driven by the SSFR: at every time, but more evidently at early epochs, 
star forming galaxies are spread around mean galaxy colours. This particularly holds true for the fainter 
(see Fig. \ref{uv}) and less massive (Fig. \ref{uvms}) systems --and is mirrored by the scatter evolution 
in the central panel of Fig. \ref{slosca}.
Groups behave differently as they show larger scatters due to their 
larger activity -- involving also more massive galaxies -- which continues also at late epochs, 
i.e. until $z\sim 0$. 

In particular, more massive galaxies in groups form a less defined RS than in clusters,
especially in the redshift interval 0.7--1.5:
the tightness of the CMr spreads up from $z\sim$1 onwards for clusters, but already at 
$\sim$0.5 for normal groups: these are the only class where the scatter is already
slightly increasing even at low $z$, whilst it keeps fairly constant in the other cases,
staying within 0.2 mag.
As for clusters, Stanford et al. (1998) had found that the 
scatter does not increase up to $z\sim 1$, Blakeslee et al. (2003) even up to $z$=1.3. 

On the other hand, quiescent galaxies (over all masses and environment)
populate a very narrow strip through all times:
its scatter mostly increases because the number of dead galaxies increases. 
Its slope is almost constant until $z$=1.5 $ (\alpha \simeq 0$, with a median value $\alpha$=-0.02) 
and at low $z$ coincides with that of RS, at least in clusters. 
Under this regard, the DS results confirmed as the locus
toward which all the galaxies asyntotically converge once quenched their star-formation, and 
seems to be a sort of universal intrinsic property of dead galaxies similar to a scale relation in 
the colour-luminosity plan. 
 
In more detail, the scatter of the DS 
is very small at early epochs and slightly increases out to $\sigma_{(U-V)}=0.15$ right after $z\sim$1,
where it is dominated by the low mass galaxies which have a larger scatter than the massive ones.
The slightly increasing trend of the DS scatter can be understood as a consequence of galaxy ageing. 
Being the DS, by definition, free of the SSFR effects 
(as built of dead galaxies), the scatter of the DS is only function of the age. 
Indeed, the spread of ages is smaller at high $z$ and gets larger at low $z$ due to 
an asynchronous formation history of galaxies (see Fig. \ref{agemass}).     
This means that either in clusters and groups there are many low mass galaxies (mostly 
responsible of the 
$\sigma_{\rm U-V}$ scatter) switching off their star formation since $z\sim1$, which are also the younger ones.


%
At the bottom of Fig. \ref{slosca} we finally plot the intercepts of the RS' fits taken at $M_V$=-20.5, 
along with the DS values as shaded region. As a comparison with observations 
data points from Bell et al. (2004) are plotted too. 
The zero-point evolution turns out to be the least affected by mass, environment and even galaxy
structural properties: the mean colour regularly increases over times from a value of 0.5 to 1.3,
also matching the observational trend (see also Holden et al.,2004). 
Galaxy cores are systematically the reddest environments,
normal groups the bluest, but not at neither very early nor very late redshifts, where all the
galaxy populations appear to both start and finish as one with same colours.

\begin{figure*}
\centering
\includegraphics[width=8cm]{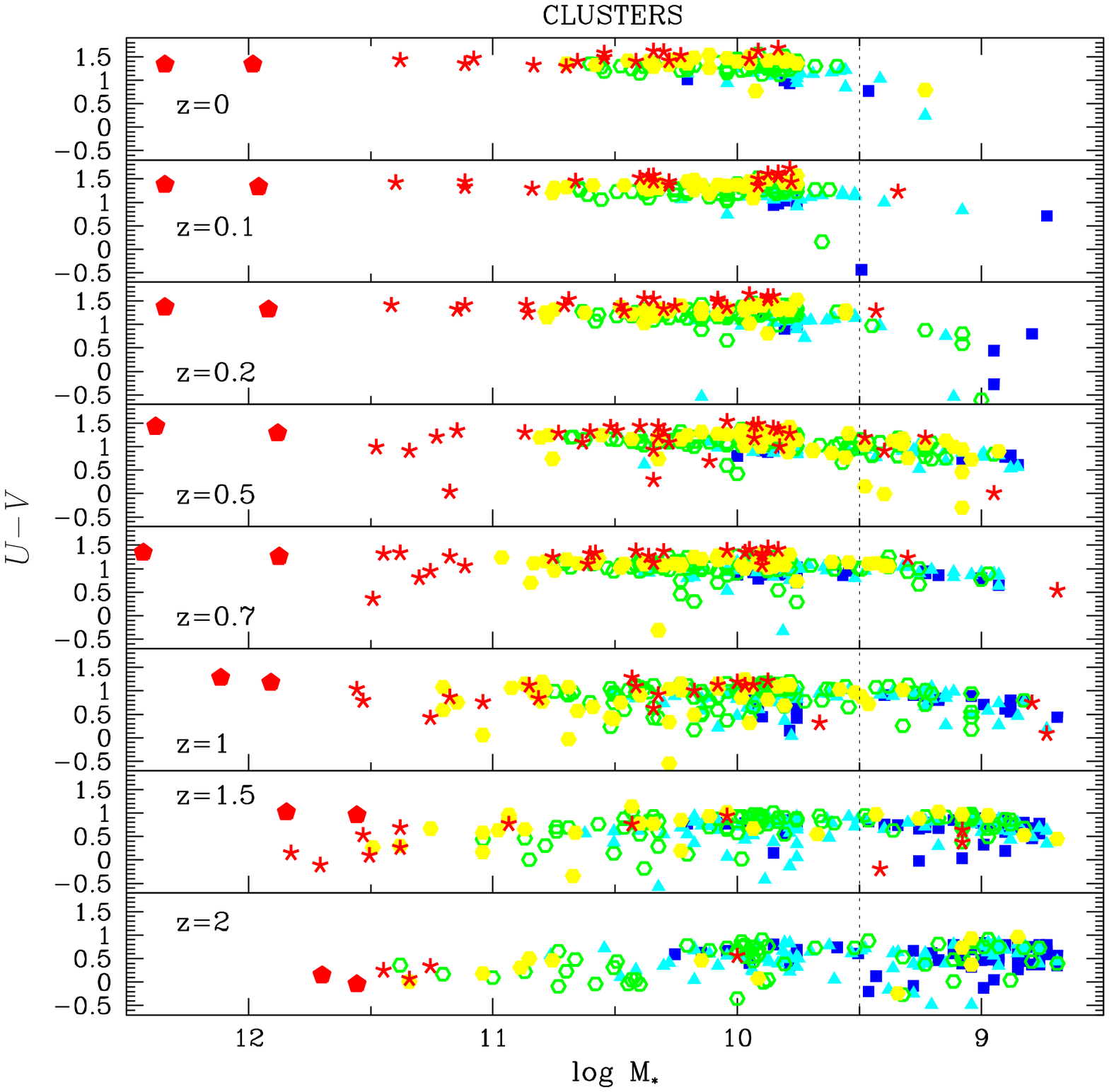}
\hfill
\includegraphics[width=8cm]{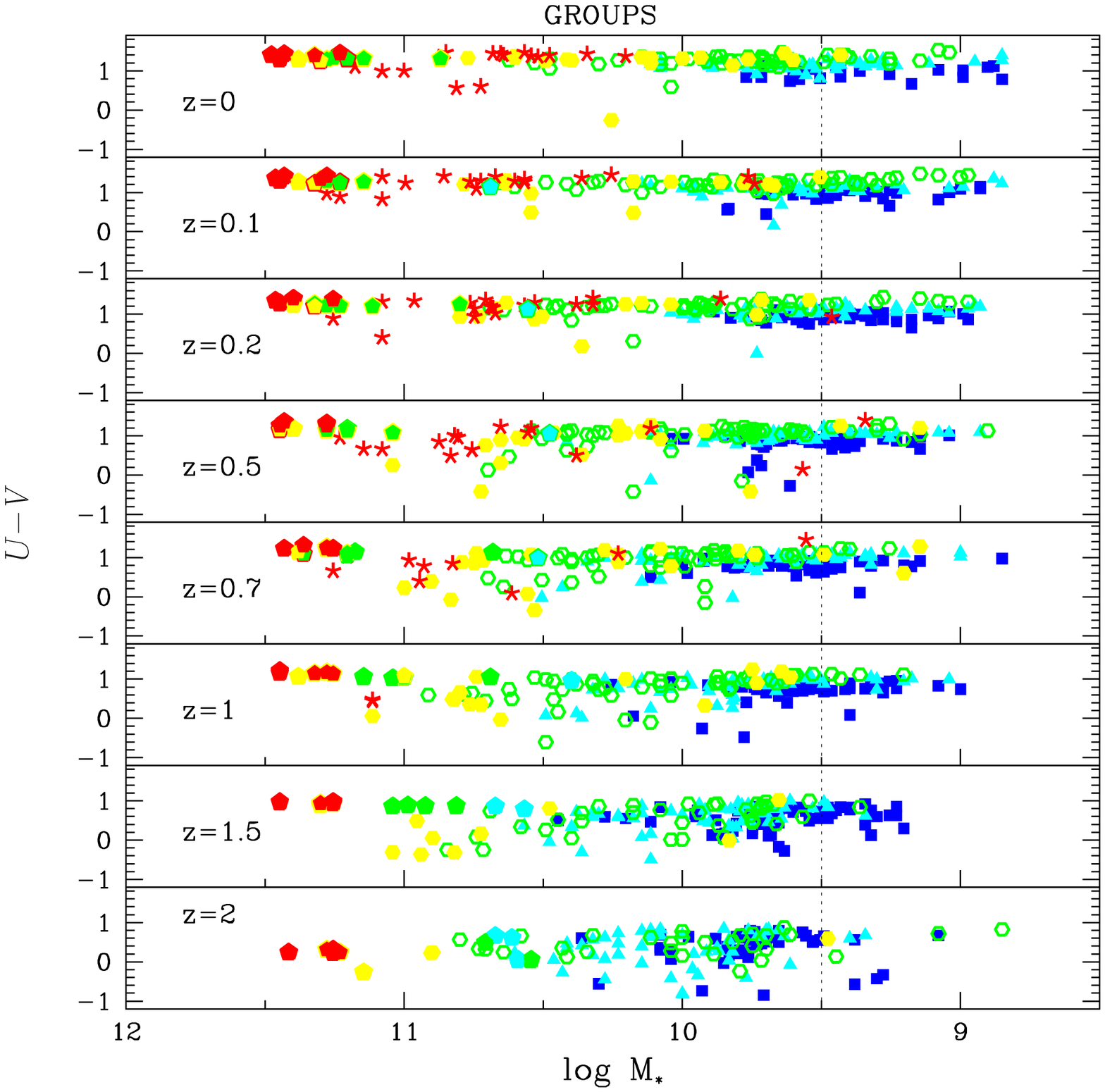}
\hfill
  \caption{{\it U-V} colour-stellar mass relation for galaxies in clusters (left)
  and in groups (right) at different redshifts. 
  Galaxies are classified by their mass-weighted metallicity, in solar units:
$Z_*/Z_\odot \leq$0.3 (blue squares), 0.3--0.5 (cyan triangles), 0.5--0.8 (green hexagons),
0.8--1.1 (yellow hexagons), $>$1.1 (red stars).
  } 
\label{uvmet}                  
\end{figure*}

\begin{figure*}
\centering
\includegraphics[width=8cm]{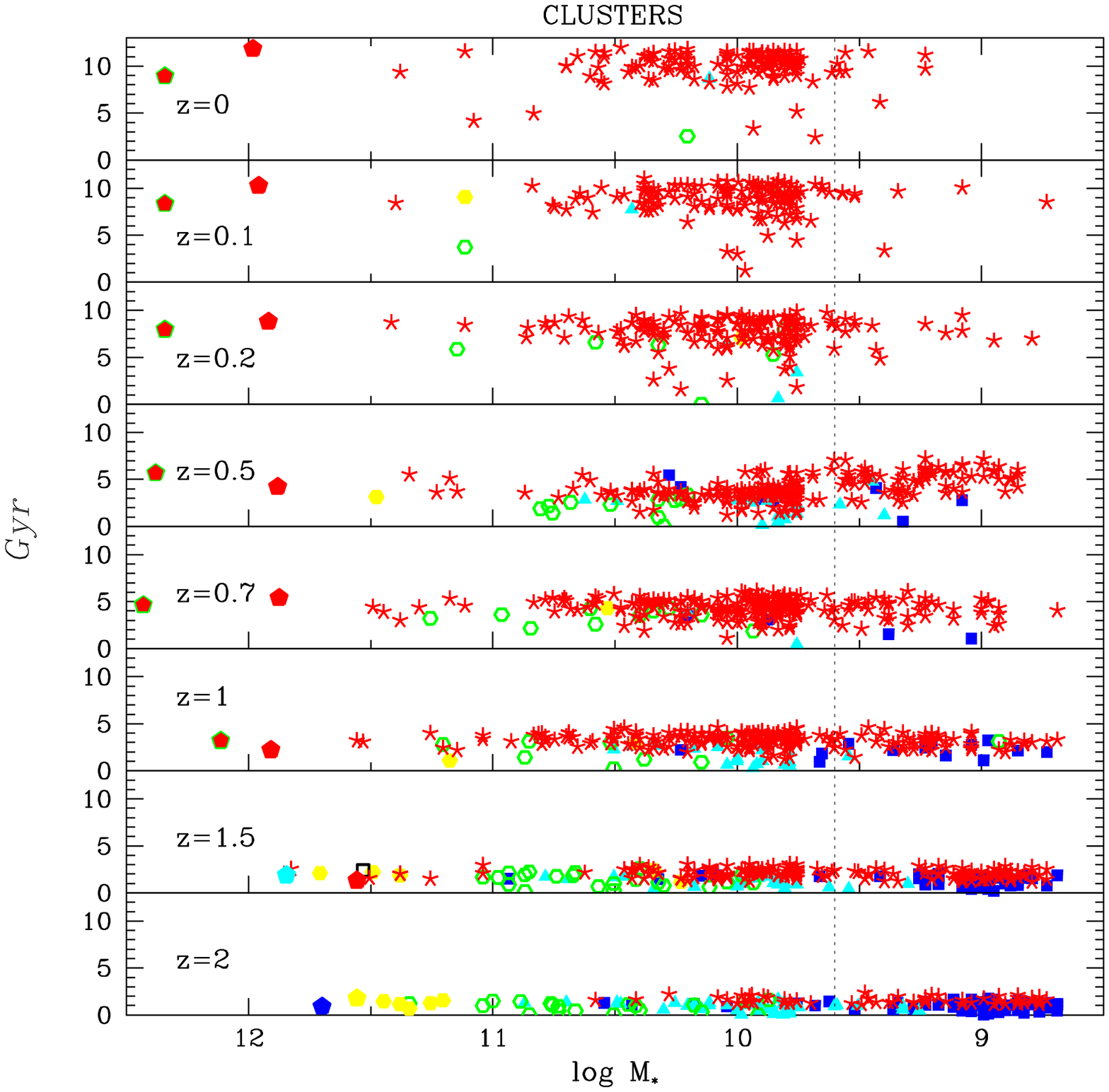}
\hfill
\includegraphics[width=8cm]{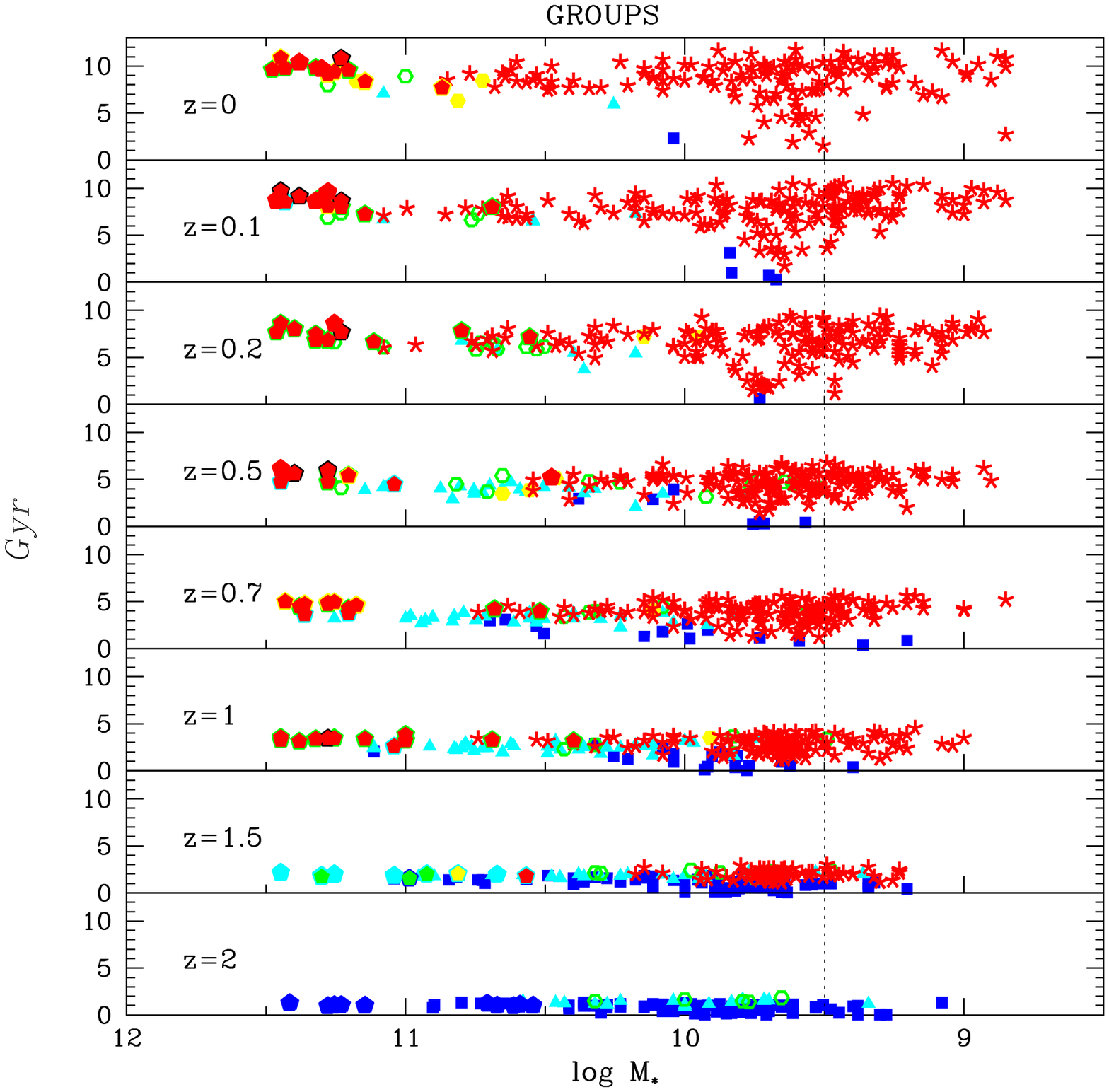}
\hfill
  \caption{Age--mass relation for galaxies in clusters (left)
  and in groups (right) at different redshifts, classified by SSFR as in Fig. \ref{uvms}
} 
\label{agemass}                  
\end{figure*}


Following the discussion made so far, we can interpret the trend of the DS zero-points 
over time as the path of the passive evolving population, as this is the sample which is not star-forming anymore. 
This is coherent with the interpretation from Kodama \& Arimoto (1997) of bright cluster galaxies being 
consistent with a passive evolving galaxy population. 
From this point of-view, looking at Fig. \ref{slosca} we can argue that galaxies in clusters start to 
mainly behave 
as passive systems much earlier than galaxies in normal groups.     

The bottom-line is that the DS appears to 
represent a quasi universal feature already established at $z\sim$1.5, after which it
does not evolve any longer, except for growing in number, according to natural passive
evolution of stellar populations. As already seen in Fig.\ref{cmobs}, the DS can thus be 
considered the ``asymptotic'' locus 
which the galaxies belonging to the RS eventually tend to, after having turned off their star 
formation activity.

\subsection{Age and metallicity as drivers of the Red Sequence}

The shape of the CMr is contributed to by metallicity and age, which both concur in determining 
the galaxy colours. Fig. \ref{uvmet} shows the colour-mass diagrams
with galaxies characterized by their metallicity (in solar units),
to be compared with Fig. \ref{uvms} where they were classified by SSFR.
In PII we had found that the CMr at $z$=0 mirrors a more general metallicity--mass relation
(as also confirmed by Gallazzi et al., 2006 and S\'anchez et al., 2007):
indeed this holds in the sense that there is a precise sequence along which more massive/luminous
objects are more metal-rich at all $z$, but this does not explain alone how galaxies pile
themselves on to the RS, which is a peculiar subregion of the colour-magnitude (or mass) plane.
The metallicity--mass relation, almost indipendent of redshift, does not directly translate in to a 
colour--mass relation: metal-richer galaxies tend to also be redder with time (see also Fig. 
\ref{colmetage} below), 
as they progressively extend toward less massive bins. 
The building of the RS rather follows such slow construction of the colour--metallicity relation,
which begins from the more massive end at around $z\sim$1 and eventually spreads up covering the same
locus populated by RS galaxies.

In Fig. \ref{uvmet} the tight correlation between mass and metallicity over times is kept
with a little scatter in metallicity at all masses: where the CMr scatter increases, this is 
not due to large variation of the galaxy 
metallicities at that mass range, particularly at the high mass side; in the low mass systems, metallicity 
can eventually add some variance to the colour distribution (see e.g. Brooks et al., 2007, for simulated field galaxies).
Metallicity can then concur in making up the RS scatter (see above) but, as we will see 
in section \ref{coldis}, its contribution is much less effective 
than the SSFR on the colour evolution of the galaxies. 
 
In Fig.\ref{agemass} the age-mass relation is shown for the same galaxy samples. 
First evidence from the plot is that the older galaxies spread across all the mass range.
In both the classes, but especially in clusters,
the older galaxies are those first placing themselves on to the DS, leaving the younger
objects behind in a diffuse ``blue cloud'': the latter though is made up of both active and,
the more with time, passive objects.
The age scatter steadily increases from higher $z$ to the present, following
the natural ageing of objects that were not born simultaneously.
Age therefore seems responsible of the increasing scatter of the DS, as seen in Fig. \ref{slosca}: 
being the DS, by definition, independent of SSFR (as made up of dead galaxies), 
its scatter is then function of the age only (and possibly metallicity). 
The larger age scatter at low redshift of galaxies populating the DS means that there must be 
young systems quenching their star-formation since $z\sim1$, mainly in the low mass regime where 
they also result as metal poorer (probably dwarf galaxies).

%

\begin{figure*}
\centering

\includegraphics[width=4cm]{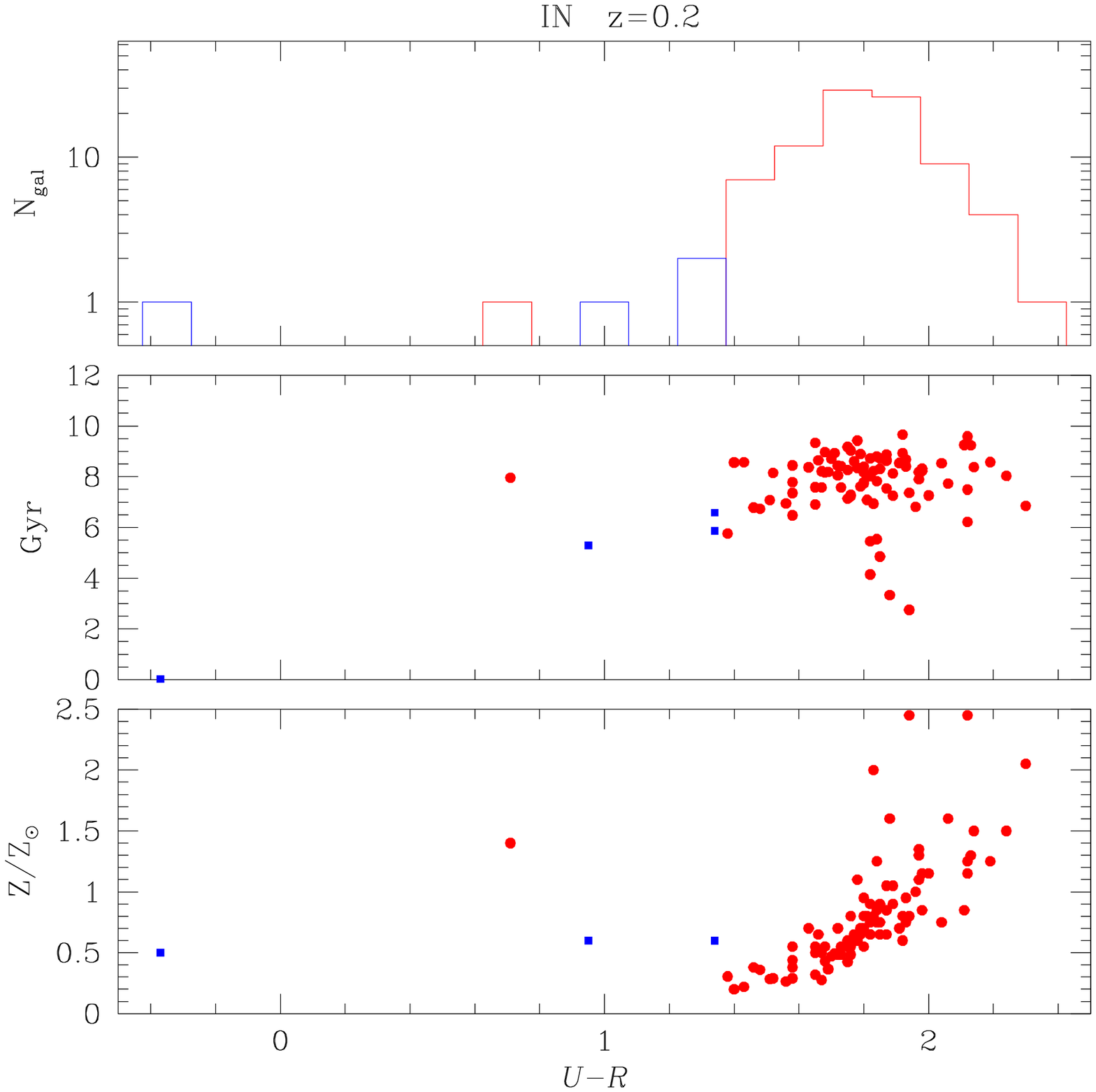}
\includegraphics[width=4cm]{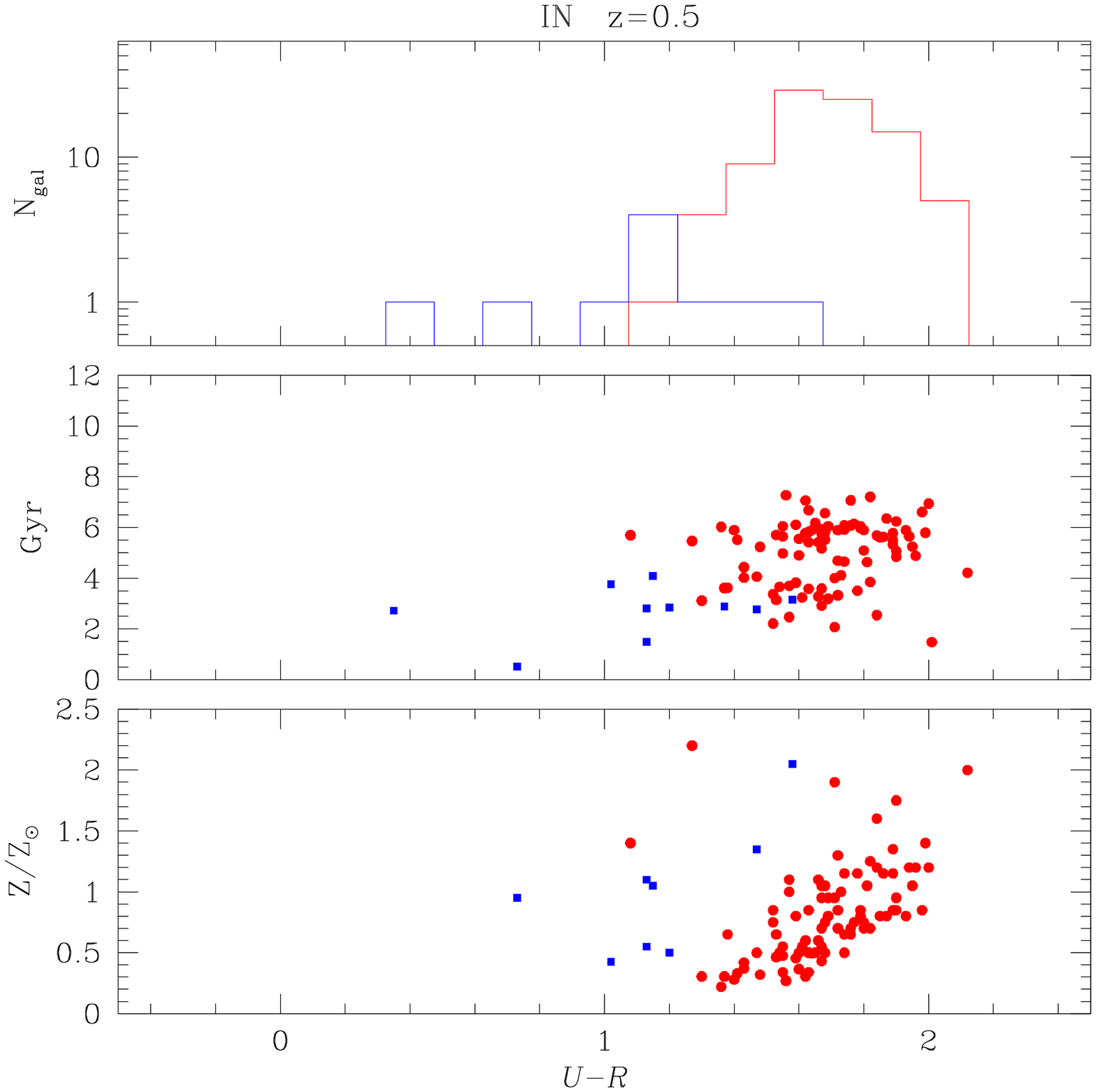}
\includegraphics[width=4cm]{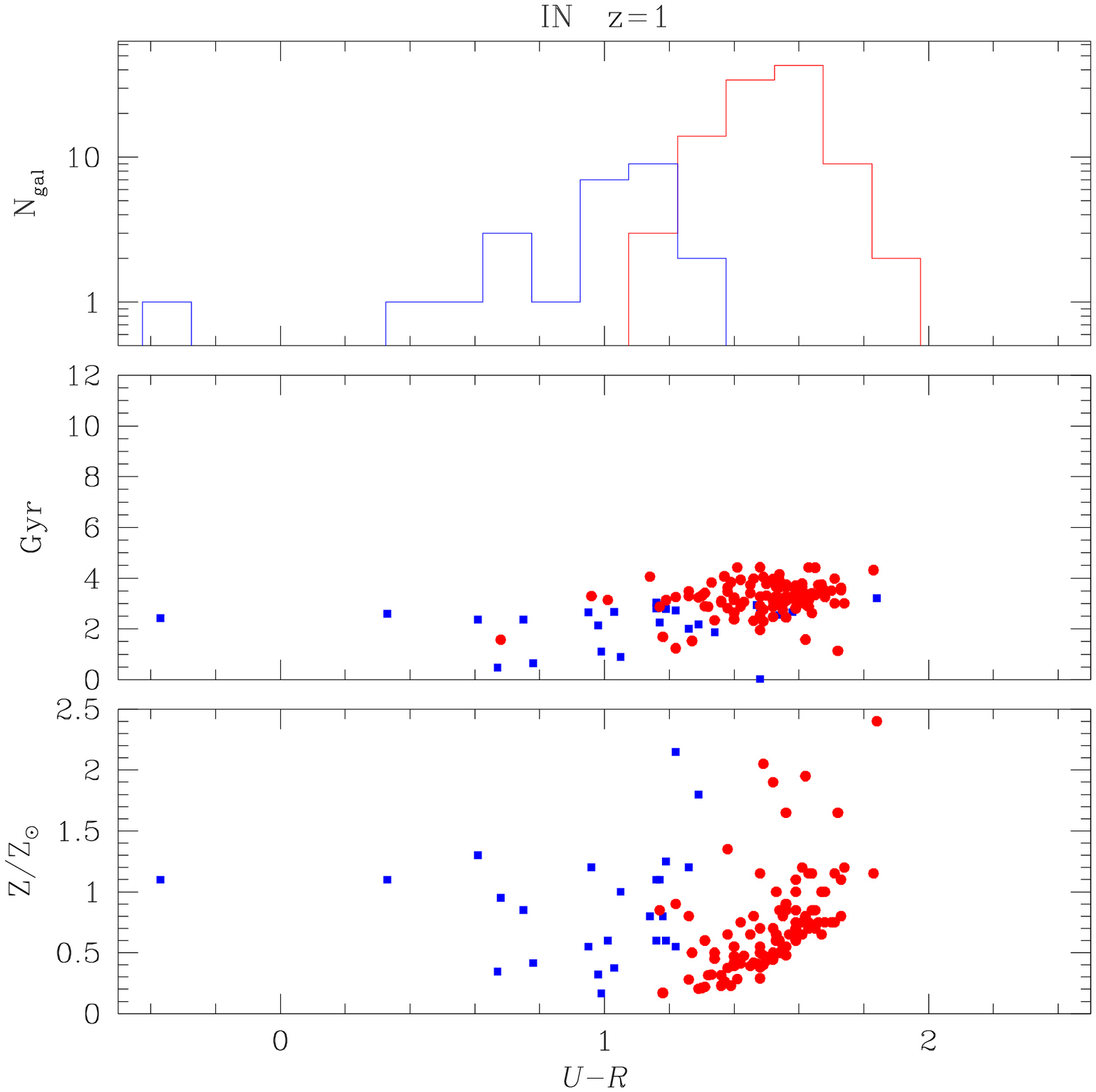}
\includegraphics[width=4cm]{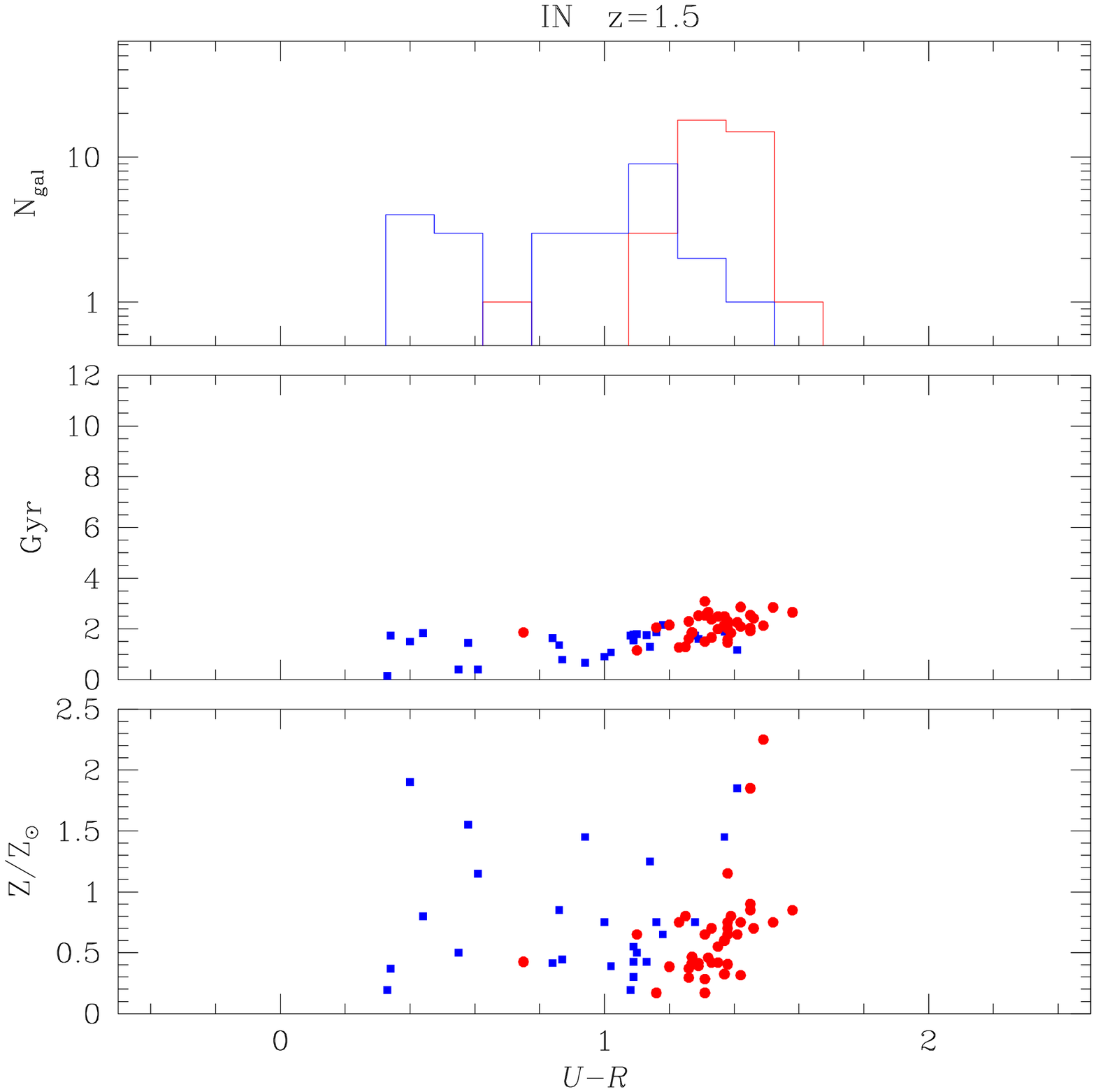}
\includegraphics[width=4cm]{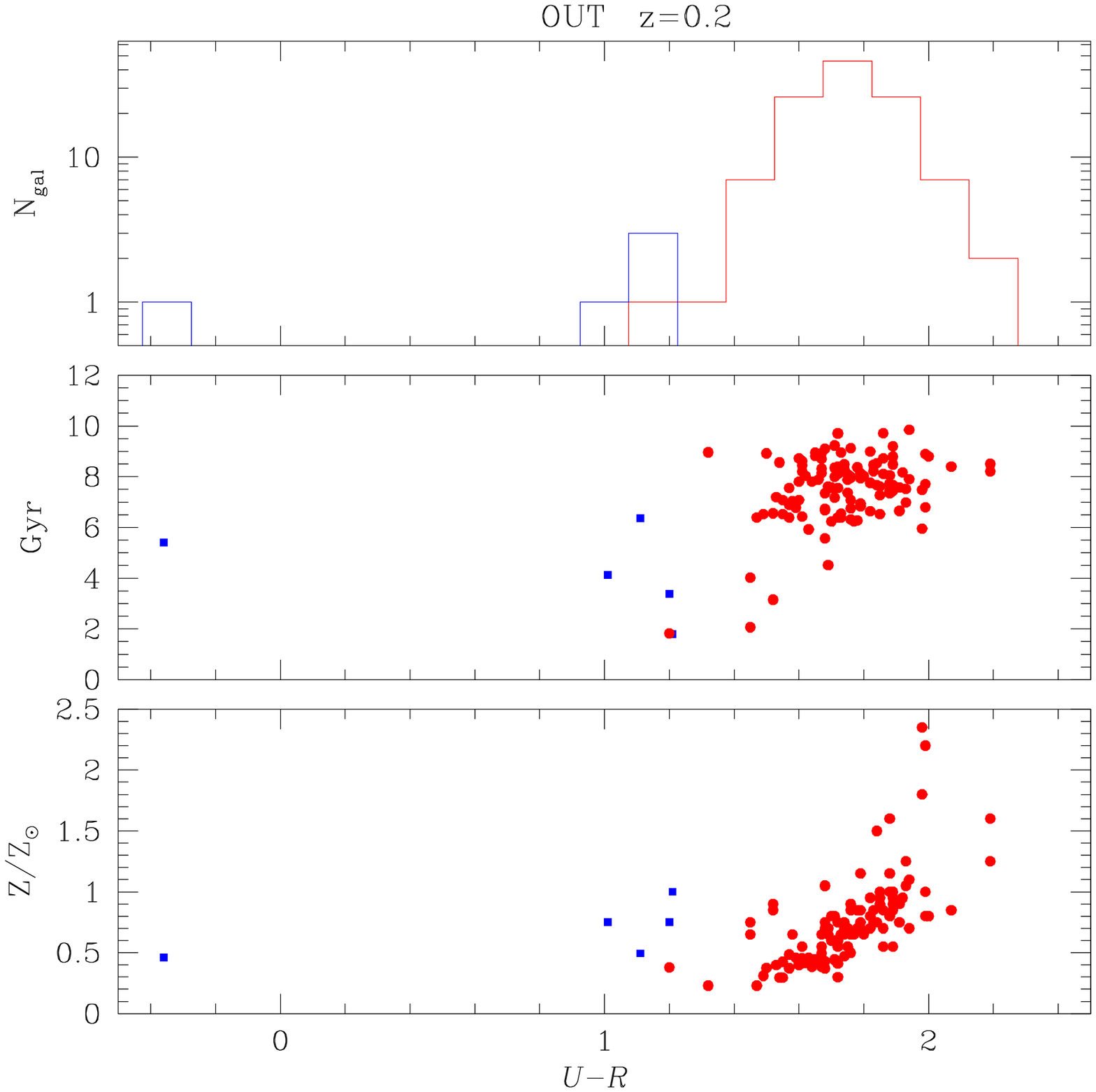}
\includegraphics[width=4cm]{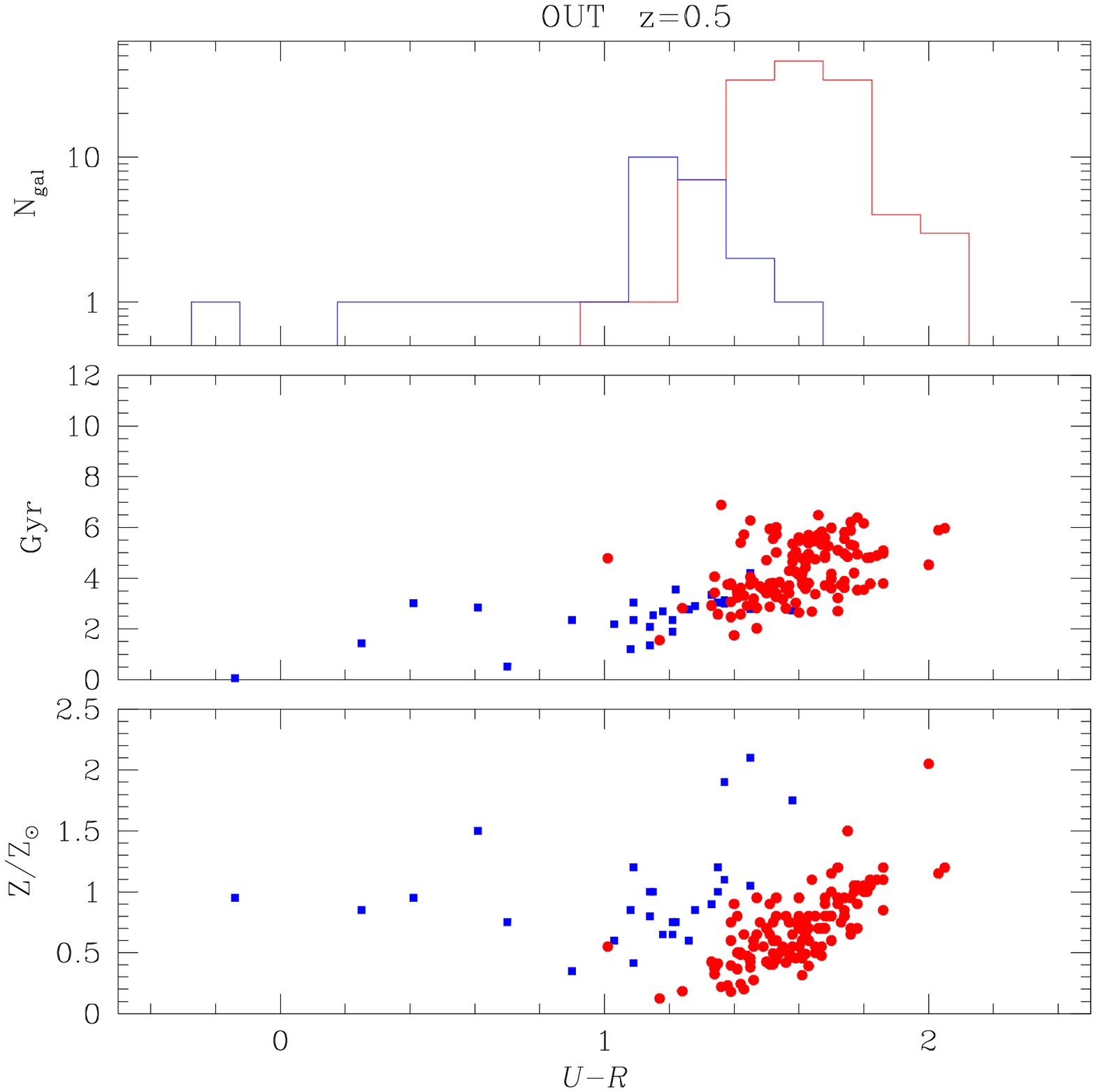}
\includegraphics[width=4cm]{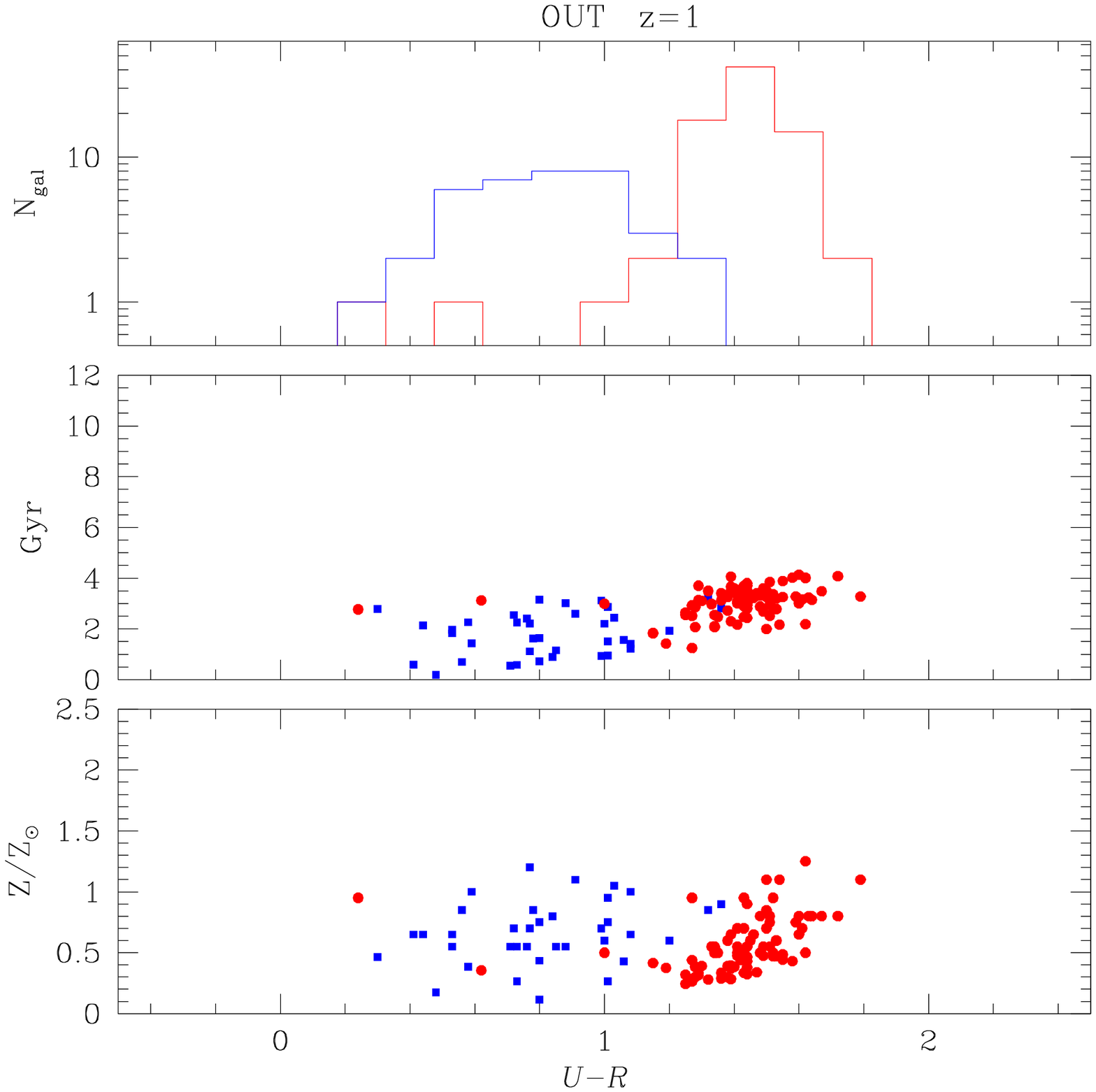}
\includegraphics[width=4cm]{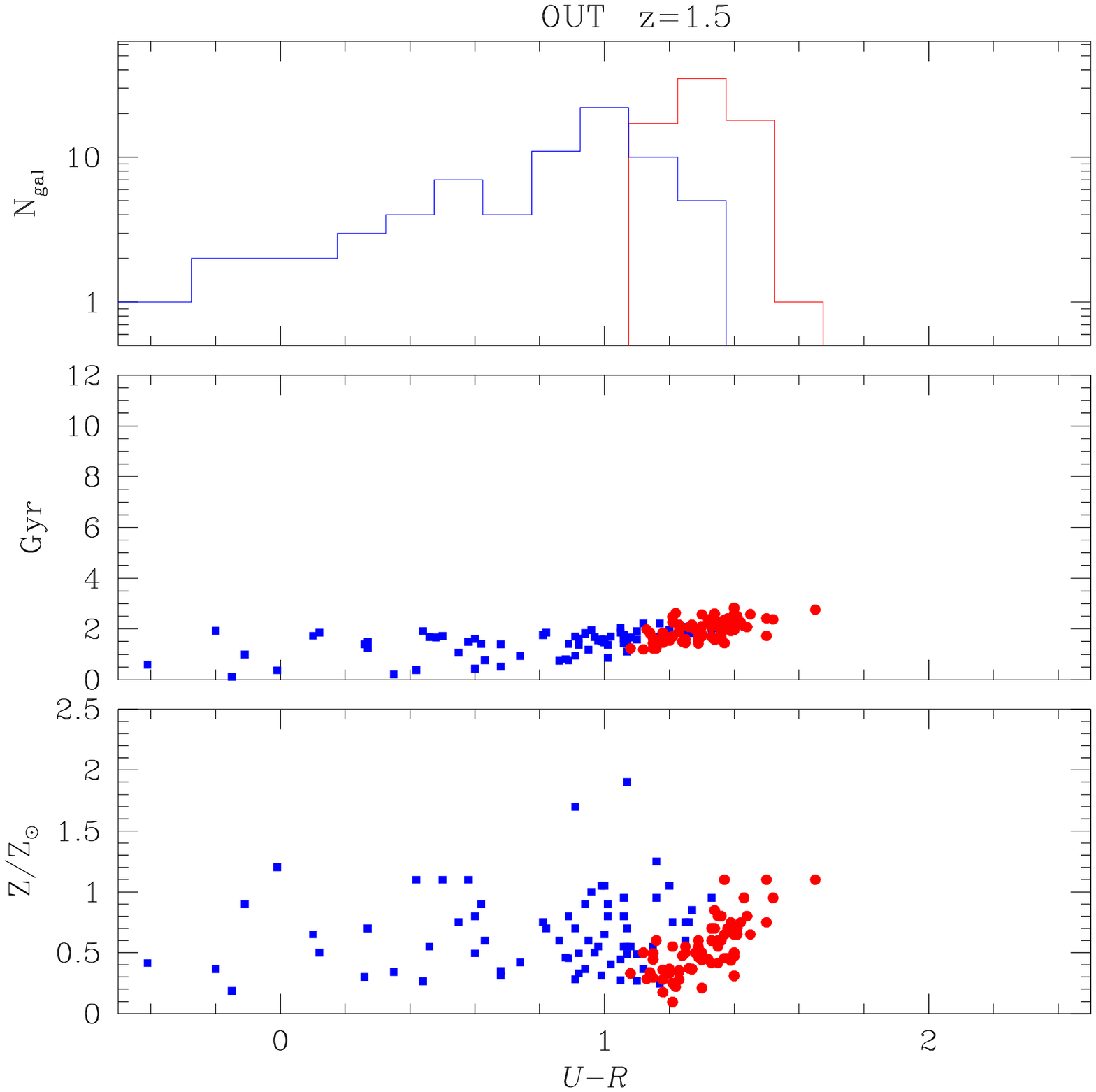}
\includegraphics[width=4cm]{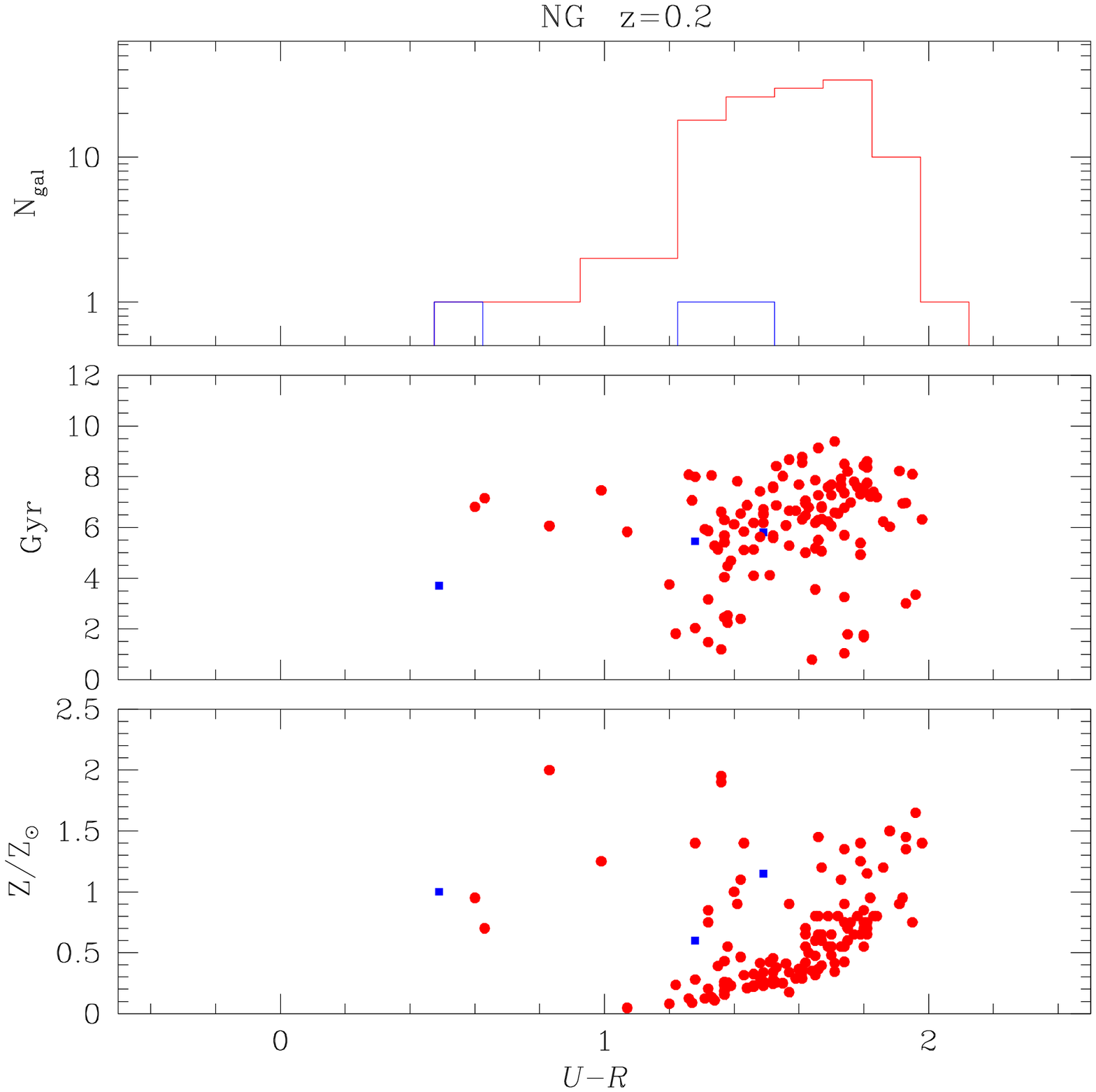}
\includegraphics[width=4cm]{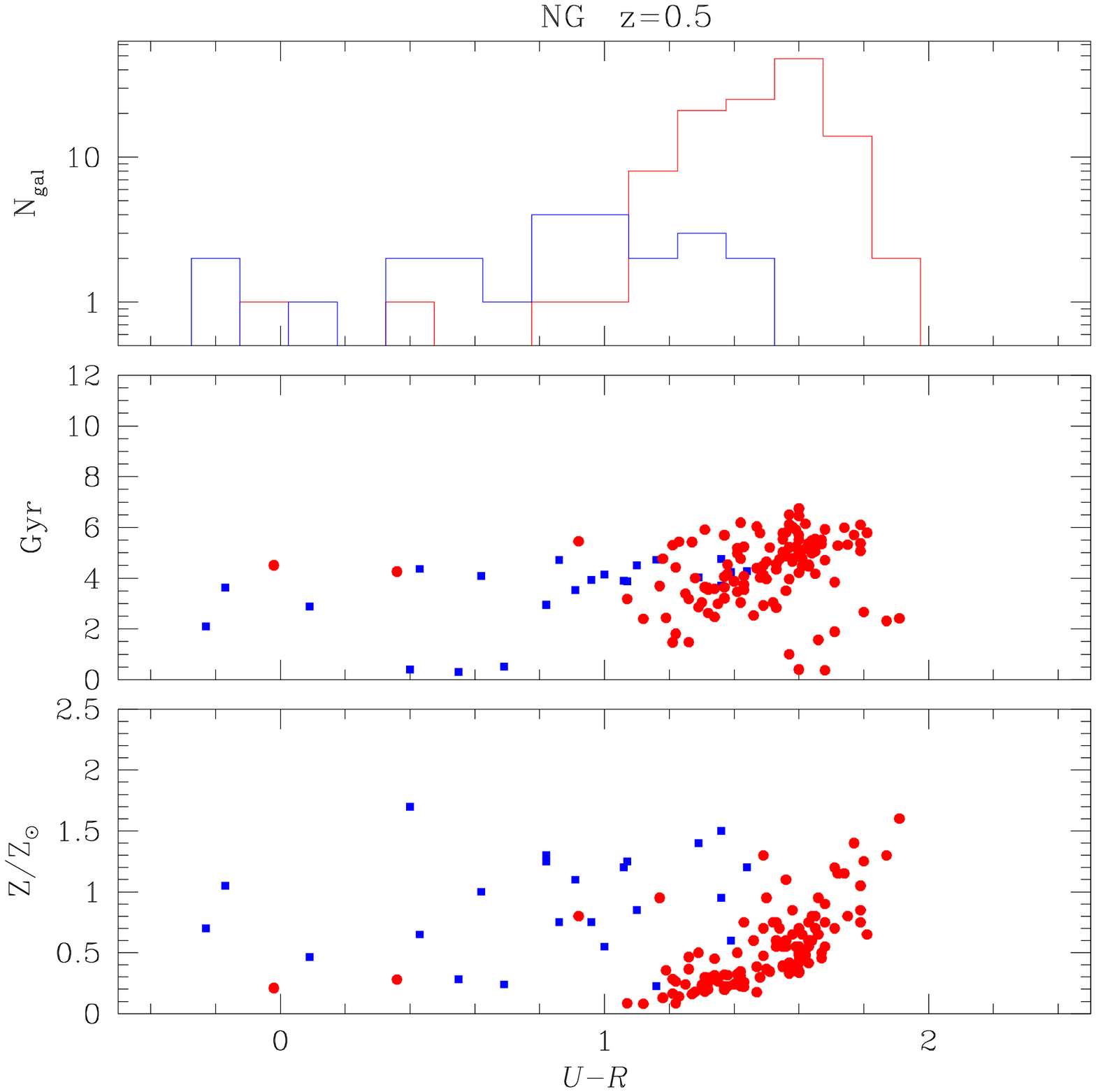}
\includegraphics[width=4cm]{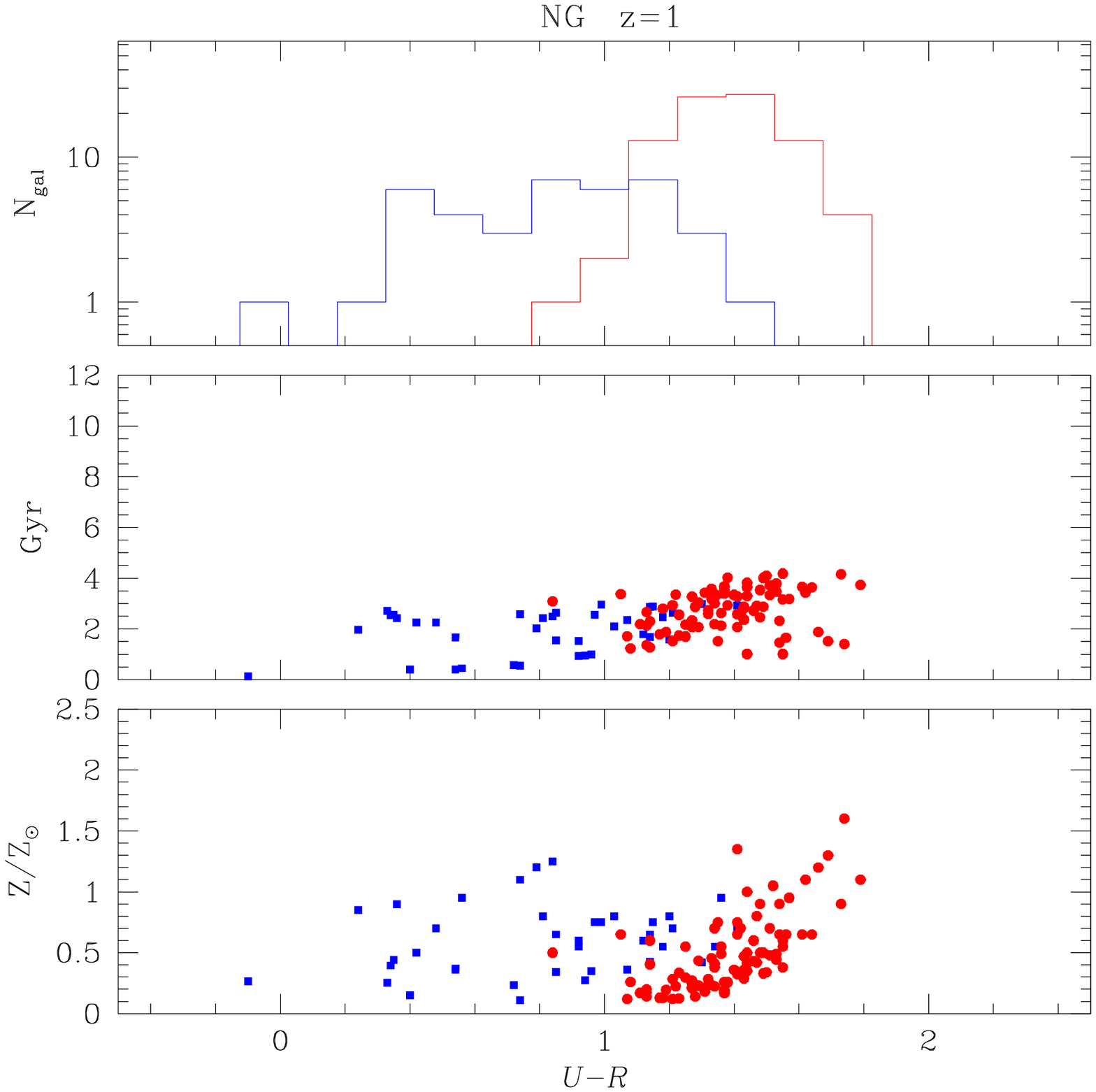}
\includegraphics[width=4cm]{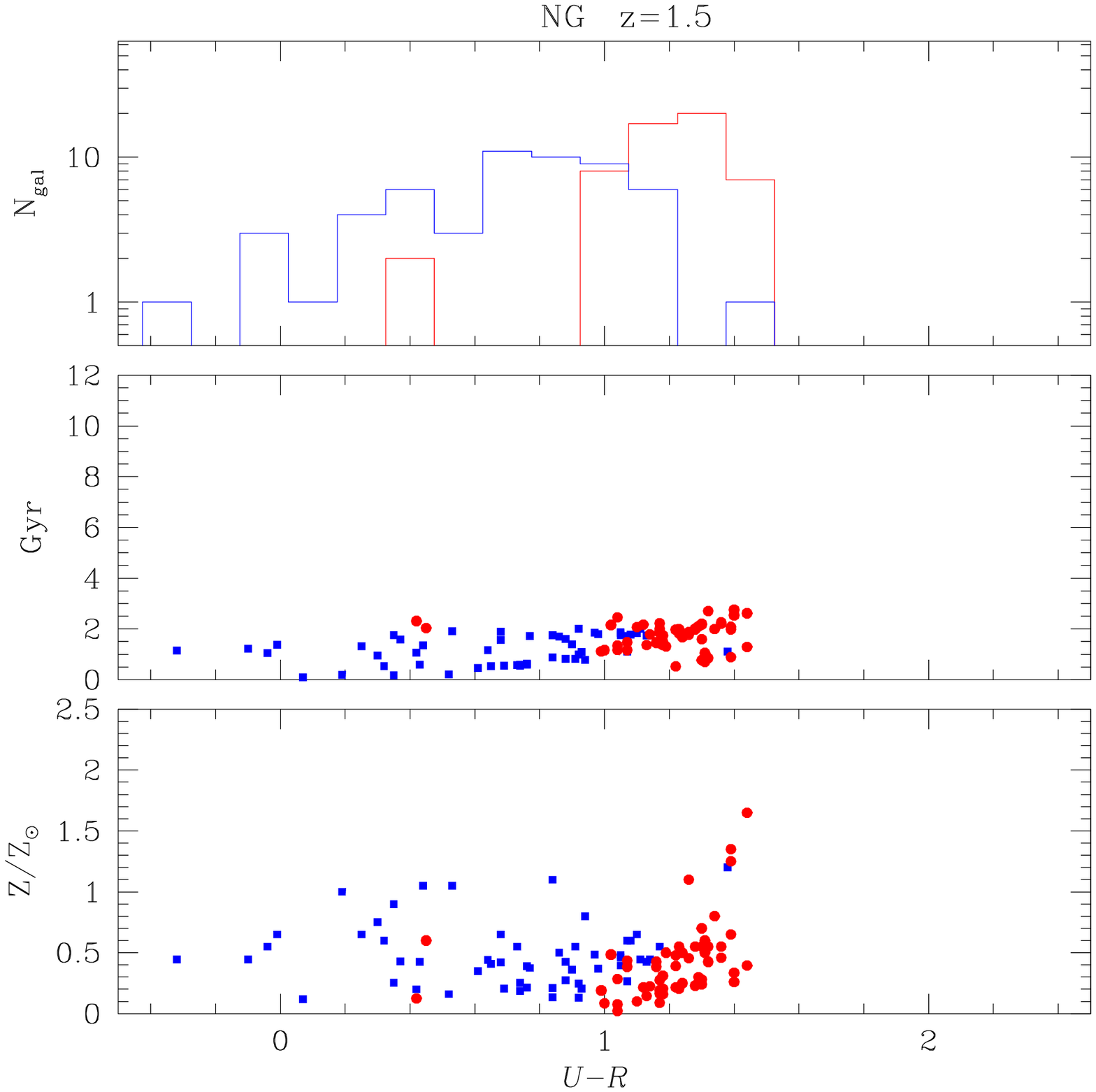}
\includegraphics[width=4cm]{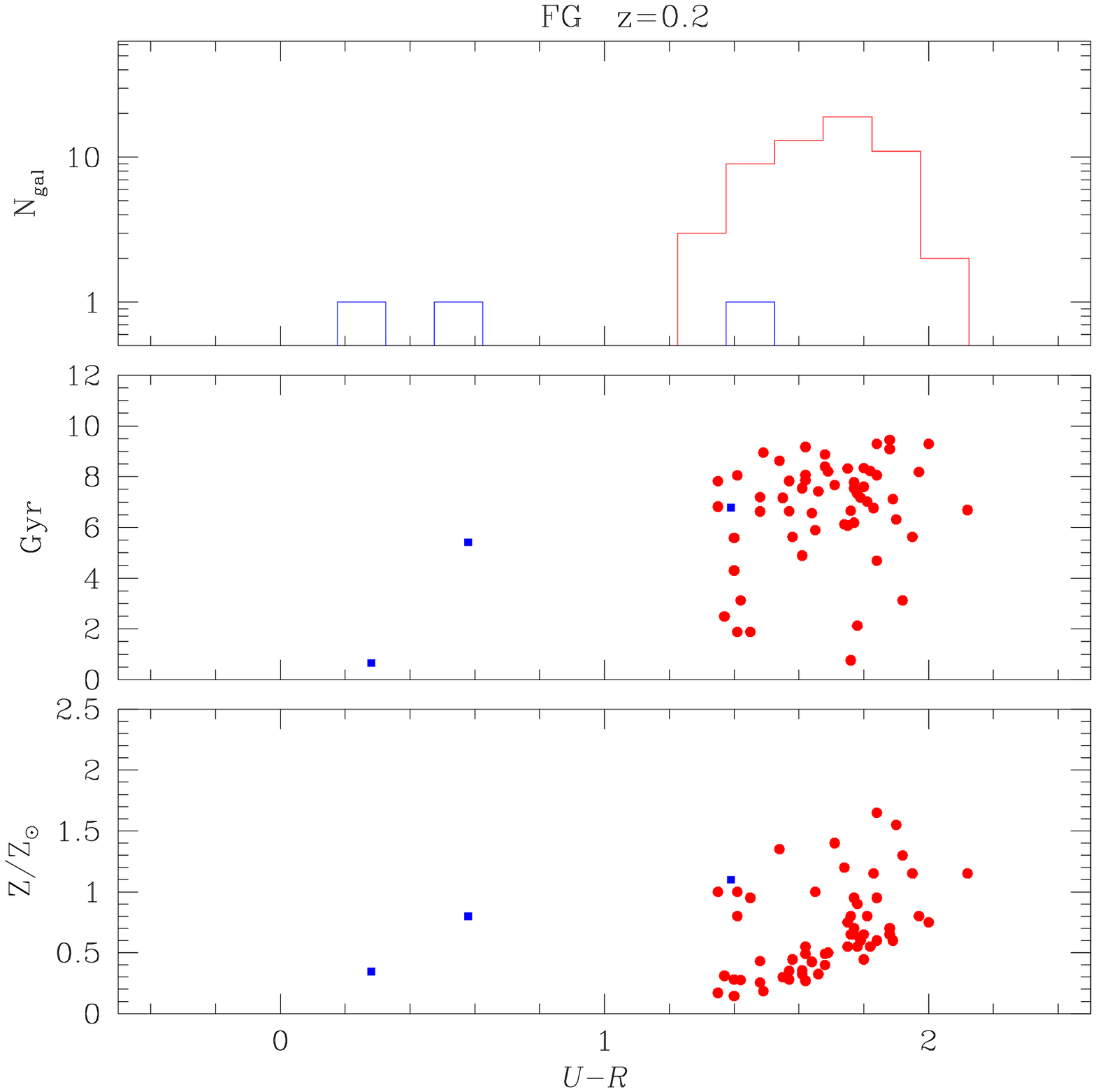}
\includegraphics[width=4cm]{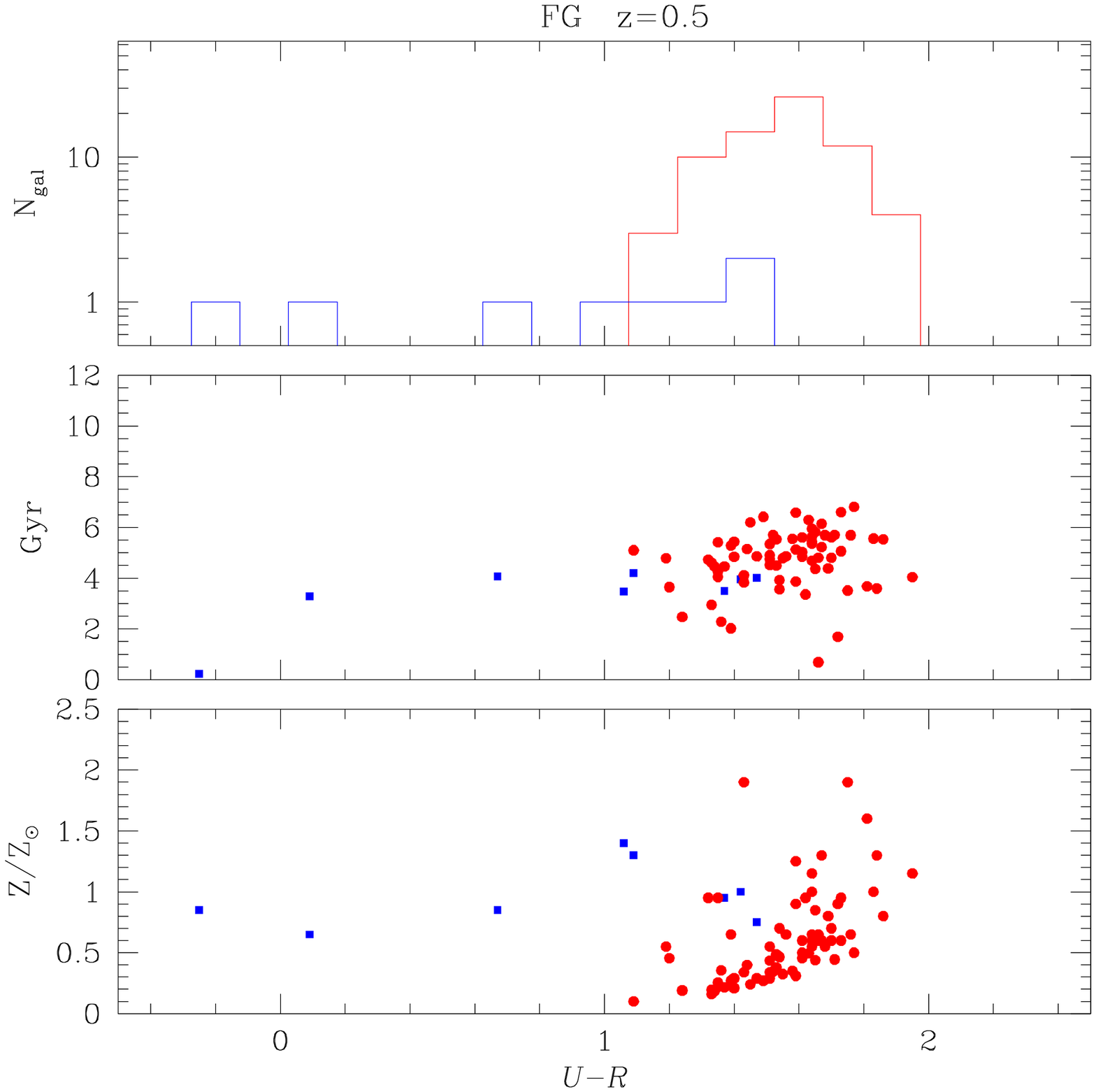}
\includegraphics[width=4cm]{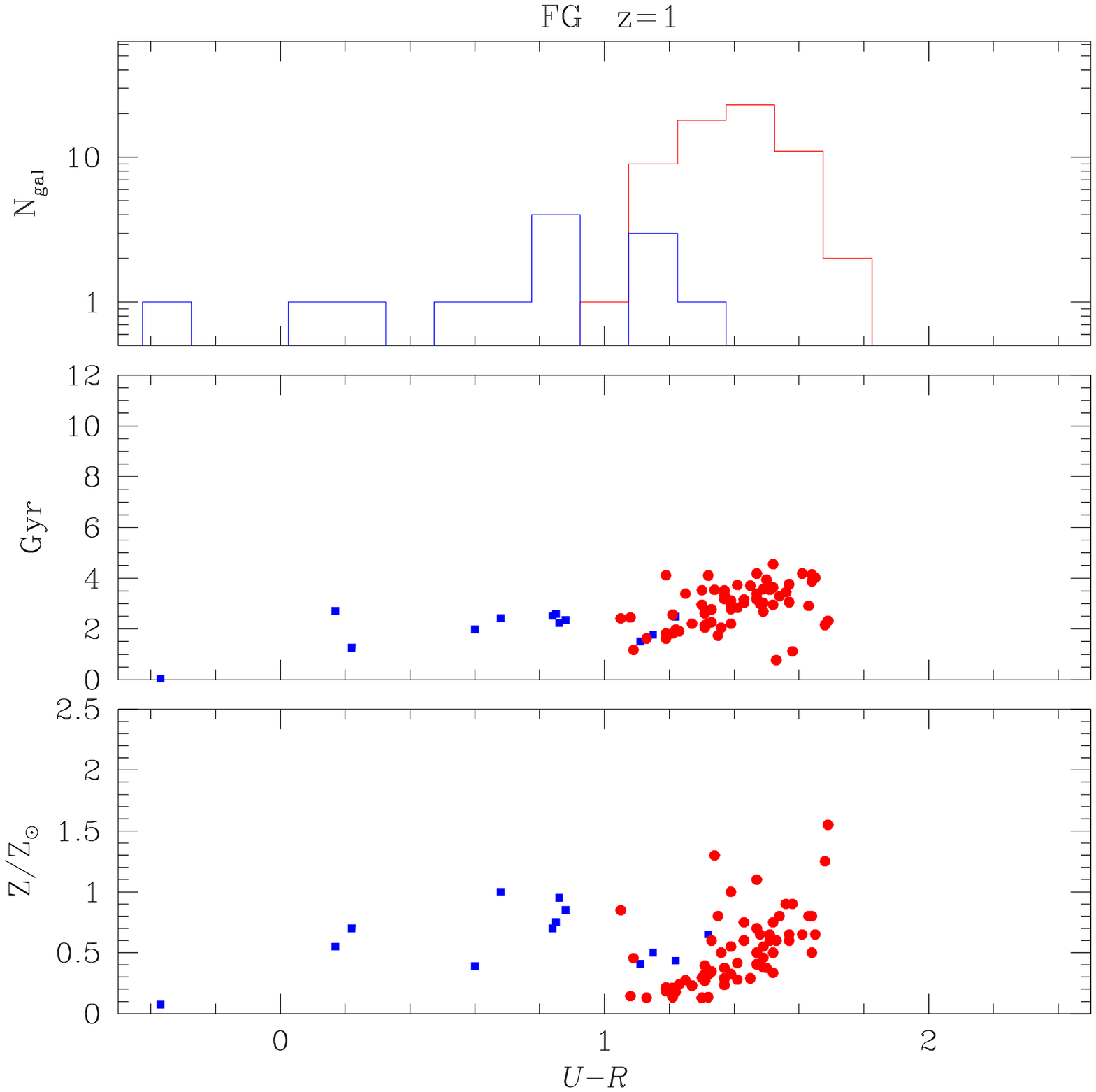}
\includegraphics[width=4cm]{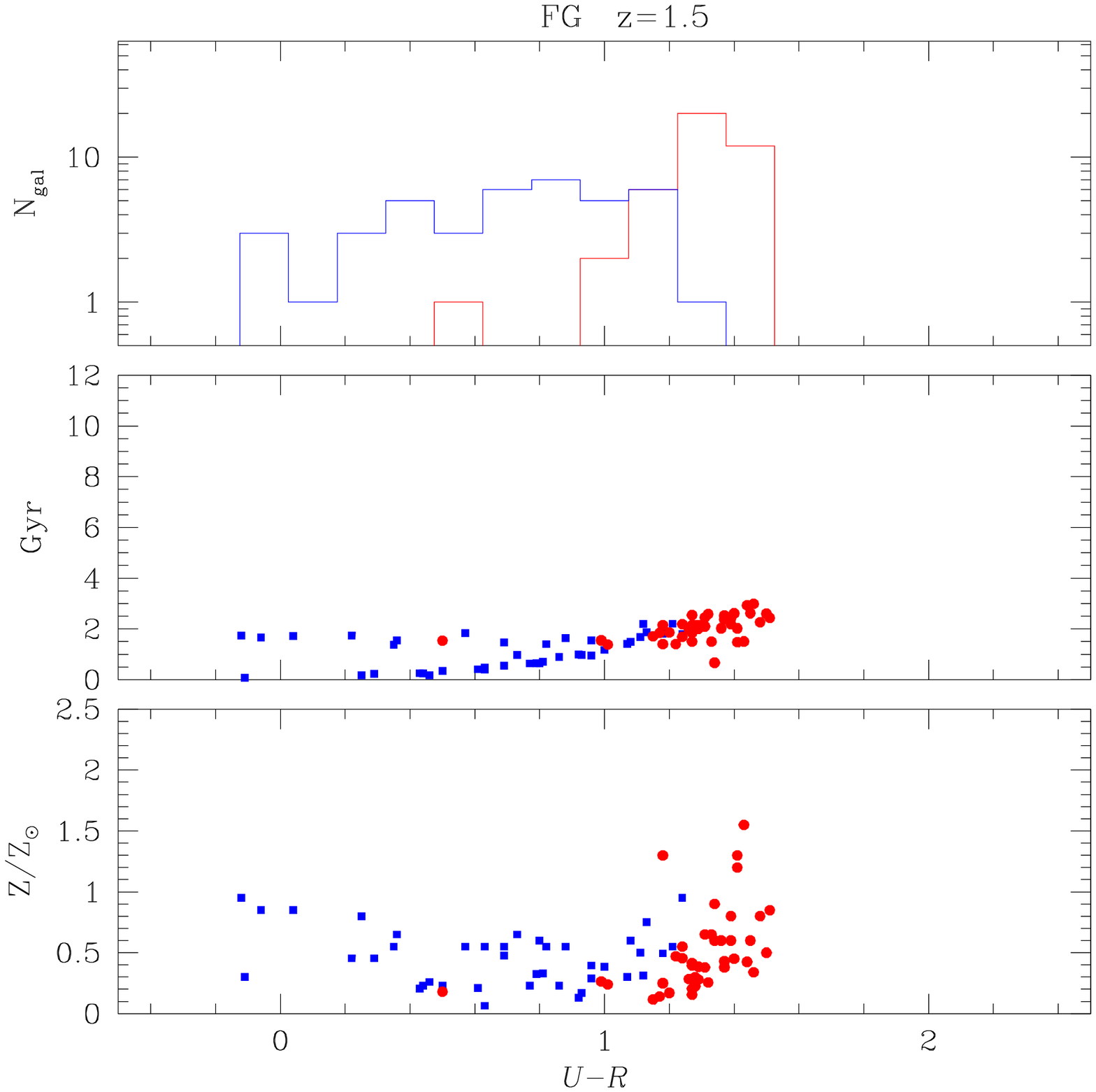}

  \caption{{\it U-R } colour distribution of galaxies, along with age--colour and metallicity--colour
relations, for galaxies
in the cluster cores (first row), cluster outskirts (second),
  normal groups (third) and fossil groups (fourth) at different epochs:
  shown respectively in blue/red are objects with specific star formation rate (SSFR) 
  during the last Gyr higher/lower than a value equivalent to 1 $M_{\odot}/yr$ for a
  $10^{10}M_{\odot}$ galaxy.}
\label{colmetage}                  
\end{figure*}

\subsection{The colour distribution}\label{coldis}

The distribution of rest-frame colours in cluster galaxies has
been found to have a rather clear bimodal shape, which translates
in to two distinct CMr, one for ET galaxies (the RS)
and one as well for late-type ones (Gavazzi et al. 1996). Such bimodality
was then confirmed in a wider range of cosmic environments and redshifts, up
even to $z\sim 2$ (Bell et al. 2004, Franzetti et al. 2007),
even though no difference due to environment was found in the loci of the two
sequences in the colour-mass relation (Baldry et al., 2006).
Like bimodality in the distribution
of SFR (Balogh et al. 2004), the bimodal colour distribution is expected
to arise from galaxy colours being optically dominated by young stellar
generations in galaxies with even small values of star birthrates. 

In Fig. \ref{colmetage} we show the {\it U-R} colour--metallicity and colour--age relations, 
along with the colour distributions at different redshifts ($z$=0.2, 0.5, 1 and 1.5), having 
separated the star-forming galaxies from the quiescent ones using the usual threshold of 
0.01 $M\odot/Gyr/M_*$. 
We see that the active versus passive separation fairly accounts 
for the colour bimodality: active and quiescent systems are segregated in two 
histograms, with the formers making up a tail, shrinking with time, at the blue side of
the passive galaxy distribution. 

The colour--metallicity relation shows a tightness increasing with time in all
the classes considered, its scatter being mostly due to the star-forming population.
Moreover, the median metallicity of the samples evolves very little with redshift,
peaking at $Z$=0.5-1 solar at all $z$ and in all environments. 
The indications here point towards a smooth colour trend making redder galaxies metal richer 
starting from $z\sim$1, as previously seen in Fig. \ref{uvmet}.

On the other hand we also see that colour bimodality cannot be explained in pure terms 
of age neither, not having reported any neat separation of RS and blue cloud by their ages. 
The scatter in age increases with time, as the red peak gets more populated by older objects.
There seems to be a smooth sequence of age and colours in the sense that older 
galaxies are tendentially redder, but the blue cloud does not exhibit a young age segregation
--rather there are many old systems having blue colours. All in all, the age--colour
relation behaves opposite to the metallicity--colour one, becoming flatter and less defined over time.

%
We can conclude then that SSFR appears as the strongest candidate as the main driver of the 
colour evolution, as well as that of the RS shaping:
following the variation of the relative number
of galaxies in the two activity histograms, we find that it clearly evolves from an 
`initial' picture
in which the blue population dominates, through an epoch of `equipartition' at around
$z\sim$ 1-1.5, corresponding to the `transition' epoch discussed earlier.
After $z\sim1$ the relative weight of the two distributions gets inverted: a transfer 
of stellar mass from the blue to the red one is found to occur by a factor of $\sim$2 
across the
interval $0<z<1$, similarly to what has been found observationally (Bell et al. 2004).


The reddening of active galaxies occurs as a progressive colour transfusion from 
a blue indefinite cloud towards a red peak and is rather 
due to the truncation of the star forming activity: 
moreover, it seems going faster in higher density environments such as 
cluster cores (first 
row of Fig. \ref{colmetage}) and fossil groups (last row, same figure).
This can be consistent with the establishment of a colour-density relation at
already $z\sim$1, found by Cooper at al. (2006: also efficient in groups) and Kauffmann et al. (2004):
in particular the latter concluded that for intermediate mass galaxies ($10^{10}-3\times 10^{10}M_{\odot}$), 
the median SSFR decreases by more than a factor of 10 as the population shifts from predominantly 
star-forming at low densities to predominantly inactive at high densities. 

More specifically, in clusters (upper two rows) the blue star-forming
objects are predominant until $\sim 1.5$ (especially in the outskirts: see also Fig. \ref{ngal}), 
after which the blue side gets more and more 
depleted with time, contributing to populate the red peak, yet maintaining a large
scatter around $U-R\simeq 1$.  The red peak appears to be rather in
place at a definite $U-R\simeq 1.6$ already at $\sim 1.5$, since only when it shifts
towards redder colour, to eventually settling down around $U-R\simeq 2$ at present epoch.

Moving to groups (lower two rows), it seems evident that the present-day colour 
distribution differs
from normal to fossil systems: while in the formers the building of the red peak from
a blue cloud follows a rather similar evolution as in clusters, galaxies in FGs do not
distribute themselves according to a bimodal shape and the few star-forming galaxies
have on average the same colours as the passive ones up to $z\simeq 0.5$.
Only at this epoch the fossilness begins to play a relevant role in discriminating
by colour the two classes: these clearly split their behaviour
at lower redshifts, where almost all the galaxies in FGs are quiescent, whereas
a higher fraction of active galaxies is found in normal groups. 
At higher redshift, the fraction of active galaxies is comparable in all groups and
is overall higher than in cluster cores at $z\simeq 2$.


\begin{figure}
\centering
\includegraphics[width=6cm]{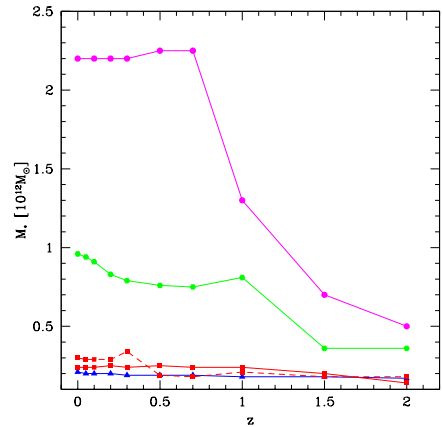}
  \caption{Stellar mass of the BCGs of clusters C1 (green hexagons) and C2 (magenta circles) and of
normal group 186 (blue triangles) and fossil groups 244 (red squares) and 189 (red squares connected by
dashed line), at different redshifts; no aperture corrections are considered here.}
\hfill
\label{cd}                  
\end{figure}

\subsection{Stellar mass assembly of the BCGs}

Along with the (few) other objects above $L^*$, the BCGs can be considered as a subsample of 
those brightest and most massive galaxies in the present Universe, which are, in
overwhelming majority, red objects. It is natural then to use this objects as a key test for the 
hierarchical formation theory.
The lower-left panel of Fig. \ref{uv} allows to follow 
the assembly of the two BCGs
of C1 and C2, especially evident in the colour-mass relation (Fig. \ref{uvms}). 
The BCG in C1
smoothly accretes mass thorough all the cluster evolutionary history (with a jump between 
$z$=1.5 and 1), while
the BCG of cluster C2, bigger and less relaxed, undergoes a major merger event 
at $z$=1, and after $z$=0.7 
it roughly doubles its mass and then settles down at almost constant value till present.
This is made clearer in Fig. \ref{cd} where the stellar mass evolution of the two BCGs is plotted 
as a function of the redshift.
Such late assembly of BCGs, at least those in clusters, goes on by means of mergers between
progenitors having old stellar populations: this picture well corroborates that of 
De Lucia \& Blaizot (2007), who find that half their final mass is locked up after $z\simeq$0.5,
typically through late, dry merger episodes involving satellites with little gas
content and already red colours.

At all redshifts up to $z\sim 1$, there are few blue galaxies bright enough to fade into the RS of the
brightest objects: this means that the most luminous galaxies on the RS either formed before
$z\sim 1$ and then kept migrating towards the bright end of the RS as fading star-forming systems (see
Faber et al., 2007), or formed through mergers after $z\sim 1$. 
This seems to be consistent with observations (Bell
et al., 2004) and can be explained within a hierarchical model in which galaxy
mergers trigger some starbursts after which the gas reservoir is depleted
(see Dekel \& Birnboim 2006), thus
truncating further star formation and letting the accreted galaxy mainly evolve by ageing;
or, in which the most luminous red galaxies form via dry mergers between gas-poor progenitors,
mechanism that is more likely to occur in higher density regions (see Khochfar and Burkert, 2003). 
In both cases the post-merger object is left with an increased luminosity, 
as one can check in Fig.1 from the jumps in magnitude at $z$=1 and 0.5 (cluster cores), 
$z$=0.3 (normal groups). 

The BCG mass evolution (see Fig. \ref{uvms}) settles down in clusters only since
$z$=0.7, after having gained mass during the whole previous period, as discussed above.
Instead it goes on smoothly in all groups since earliest times, with no major episodes
of mass accretion undergoing.
In Fig. \ref{cd} three groups in particular are shown, one normal (blue), one fossil (red) and
a third just in between, having $\Delta M_R$=2 (red with dashed line).
This allows one to follow the assembly process of the BCGs 
with respect to the `fossilness' of the group: given that the mass evolution in
groups proceeds much evenly than in the two clusters, there are still some differences even when
starting from very close ``initial'' conditions. 
Whereas in the normal group the BCG steadily and slowly grows by mass with a
constant rate at every $z$, with a final mass only 1.1 times larger than at $z$=2, in the FGs 
a major episode of mass accretion occurs at some epoch: slower at the beginning (between $z$=2 and 1.5) 
for the most fossil and steep and much later ($z$=0.3) for the almost fossil one, which had started
with the same mass as the normal group; in both cases the subsequent evolution settles down at
a quasi-constant rate, both ending with a final mass 1.7 times as much as at $z$=2.
Combining this with information from Fig. \ref{uv}, we can thus explain such behaviour with 
different star formation histories:
the activity lasts longer and the number of star forming galaxies remains higher in normal groups 
than in fossil ones, supporting the picture of the latters being rather ``quiescent", earlier assembled 
systems; in FGs the gap between the 
BCG and the 2nd brightest galaxy opens on average already since $z\sim$1.5 and then widens out.


\section{Conclusions}

The cosmological-hydrodynamic simulations we have performed provide a tool for interpretating
the formation scenario of spheroidal-like galaxies in various systems such as galaxy clusters 
and groups. In particular we have followed the build-up of the colour-magnitude
diagram, leaving the analysis of the luminosity and mass function to a forthcoming paper. 

Evolving galaxies have been traced
according to their specific star formation rate (SSFR), as they move towards a ``dead'' or ``quiet''
sequence 
they set upon once they have almost stopped the bulk of their star formation
activity. The completion of the RS along its approaching towards such a sequence
occurs in a similar way for galaxies belonging to the four environmental classes considered.
Fainter galaxies in clusters keep having significant star formation
out to very recent epochs and distribute more broadly around the RS, with respect to
the brighter objects. In groups the star forming objects, though still all lying below
the RS as in clusters, are more distributed across a wider mass range.
Besides this, the role of environment is evident when looking at cluster outskirts and normal groups,
where star formation activity holds longer than in central cluster regions.
However galaxies
undergoing infall from the outskirts to the centre keep star
formation on until they settle on to the same RS of the core galaxies.

The SSFR arises from this picture as the quantity most determinant in driving 
the shape of the RS in all the environments. The RS itself, operatively defined on to
the ET subsample at every $z$, turns out to be an
evolving feature of the colour-magnitude plane: at low $z$ it eventually tends to 
coincide with the sequence of dead galaxies, whose slope is instead fairly constant
over times.

The study of the evolution of the galaxy number density and of the colour
distribution indicates the existence of a dual population: a ``blue cloud''
of star-forming objects, predominant until $z\simeq$1-1.5, progressively
depletes itself by feeding the quiescent red peak: the latter is consolidated since
$z\simeq$1.5, as star formation activity shuts off, starting from less
massive systems. This period can then be considered the very epoch of
establishment of the RS, as the locus of eventual displacement of all
the ``dying'' galaxies. 

We also find that the mass--metallicity relation cannot explain alone the
establishment of the RS: being more massive galaxies also those metal-richer at all $z$, 
the formation of the RS proceeds together with the completion of the colour--metallicity sequence 
which starts from the massive end at an epoch, again, between $z\sim$1.5 and 1.

The time sequence along which galaxies in the various environments place themselves upon the RS
closely mirrors that regulating the relative predominance of the active over the quiescent populations:
this is found to range from $z\simeq$1.2 for cluster cores to 0.55 for FGs, down to 0.45 for normal groups.
The latters appear then as systems still undergoing bursts of star formation even at recent epochs.

This scenario seems to be confirmed by recent observational work on combined SDSS-DEEP2 
samples by Cooper et al. (2007), who interestingly observe a transition in colour 
at a redshift z$\approx$1, compatible with our findings. 
Also recent semi-analytical approaches give results consistent with those found 
in our paper using direct N-body/SPH simulations (Kaviraj et al. 2007, Khochfar and Ostriker 2008, Menci et al. 2008).



The results emerging from this paper are that we are able to reproduce
the photometric properties of galaxies 
as a consequence of the star formation activity during dark matter halos
assembly.
In this way the initial cosmological $\Lambda$CDM simulation, in which DM haloes
grew by hierarchical infall, has produced, through the interplay with the gas and
stellar components, a galaxy population for which the main driver of evolution is
given by the intrinsic ageing of the SSPs. Galaxies where a prolongued activity across
various minor star formation episodes is still occurring at low $z$, are those still not
placed on a defined RS.
A merging scheme appears to be dominant in shaping the assemblement of BCGs
in dense systems not yet dynamically relaxed: the cluster BCGs keep on accreting satellites even 
down to $z\simeq$0.5. Apart from this, for the bulk of the galaxy population
the formation process, namely the aggregation of dark and baryonic matter in a
defined gravitational well, occurs much before the star formation history gets completed.

New simulations including galaxy samples belonging to the field are currently
under progress and will allow a more complete analysis of the dependence of
galaxy properties on the environment. Moreover, the issue of AGN feedback is also
under process of being implemented in the hydrodynamical schemes, in order to overcome
the drawbacks related to the empirical treatment of cooling in the BCGs.


\section*{Acknowledgments}

This work has been realised within the project led by N.R.N. funded from CORDIS through the
FP6-European Reintegration Grant MERG-CT-2005-071440. 

The TreeSPH simulations were performed on the SGI Itanium II facility provided by
DCSC. The Dark Cosmology Centre is funded by the DNRF.

The work of V.A.-D. has been supported by the European Commission,
under the FP6 for Research \& Development, Action
{\em Transfer of Knowledge} contract
MTKD-CT-002995 .

The authors want to thank J. Blakeslee, M. Tanaka and Ch. Wolf for providing data tables of their observed 
galaxy samples to be compared with simulations; Laura Portinari for her precious collaboration, and Antonio
Pipino for fruitful discussions.

\bibliographystyle{aa}

\end{document}